\documentclass[11pt,aps,showpacs,tightenlines,nofootinbib]{revtex4}
\usepackage{amsmath,amscd,amssymb}
\usepackage[usenames,dvipsnames]{color}
\usepackage{xfrac}
\usepackage{framed}
\usepackage{ulem}
\usepackage{latexsym}
\usepackage{mathrsfs}
\usepackage{graphicx}
\usepackage{bbm}      
\def\beq{\begin{equation}}
\def\eeq{\end{equation}}
\def\be{\begin{eqnarray}}
\def\ee{\end{eqnarray}}

\newcommand{\Vek}[1]{\mbox{\boldmath$#1$\unboldmath}}

\newcommand{\vek}[1]{\mathbf{#1}}

\newcommand{\uu}{\hat{u}}
\newcommand{\EE}{\hat{E}}
\newcommand{\RR}{\hat{R}}

\newcommand{\deriv}{\widetilde{\nabla}}

\newcommand{\td}{\hat{t}}
\newcommand{\Field}{\Theta}

\begin{document}

\title{Quantum Stabilization of a Closed Nielsen--Olesen String}

\author{M. Quandt$^{a)}$, N. Graham$^{b)}$, H. Weigel$^{c)}$}

\affiliation{
$^{a)}$Institute for Theoretical Physics, T\"ubingen University
D--72076 T\"ubingen, Germany\\
$^{b)}$Department of Physics, Middlebury College
Middlebury, VT 05753, USA\\
$^{c)}$Physics Department, Stellenbosch University,
Matieland 7602, South Africa}

\begin{abstract}
We revisit the classical instability of closed strings in the Abelian Higgs 
model. The instability is expressed by a vanishing energy as the
torus-like configuration shrinks to zero size. We treat the radius of
the torus as a collective coordinate and demonstrate that a 
quantum mechanical treatment of this  coordinate leads to a
stabilization of the closed string at small radii.
\end{abstract}

\maketitle

\section{Introduction and Motivation}
\label{sec:IM}

The study of string or vortex type configurations in the Standard Model and related 
theories has attracted much attention in the recent 
past~\cite{Hindmarsh:1994re,Kibble:1976sj,Vilenkin:1994,Achucarro:2008fn,Copeland:2009ga}.
Because these configurations may have cosmological implications and to 
distinguish them from the fundamental objects in string theory,
they are commonly called \emph{cosmic strings}. The most imperative question regarding 
the dynamics of a cosmic string is its stability. Studies of this question 
have mainly focused on infinitely long and axially symmetric string 
configurations~\cite{Nielsen:1973cs,Nambu:1977ag,
Vachaspati:1992fi,Naculich:1995cb,Achucarro:1999it,Groves:1999ks,
Stojkovic,Bordag:2003at,Graham:2004jb,Graham:2006qt,Schroder:2007xk,
Baacke:2008sq,Weigel:2009wi,Weigel:2010pf,Lilley:2010av,Weigel:2010zk,
Graham:2011fw}. Unless these configurations carry (topological) charges,
as is the case,
for example, for the Abelian Nielsen-Olesen string~\cite{Nielsen:1973cs},
they are classically unstable but can eventually be stabilized by quantum
effects. One such possibility would be to bind (a large number of)
fermions at the center of the string~\cite{Weigel:2010zk,Graham:2011fw}.
Other stabilizing mechanisms involve the coupling of external fields 
or currents leading to superconducting strings~\cite{Witten:1984eb}.

Strings at the boundary of regions with different vacuum expectation
values are expected to build a network, so that they have the
potential to form cosmic string loops
\cite{Witten:1984eb,Davis:1988ij}. As a consequence, it  is
conceivable that closed strings or a network thereof existed at
early times of the universe when the vacuum manifold  was not simply
connected. {Cosmologically motivated studies on networks of 
closed strings suggested their stability (against graviational 
radiation) for radii of the order of a a small fraction 
$\sim \mathscr{O}(0.01\ldots0.1)$ 
of the horizon distance. These studies started from analytic 
investigations~\cite{Vilenkin:1981iu,Kibble:1982cb,Burden:1985md}
but were followed by intensive numerical simulations, cf. the
recent article~\cite{BlancoPillado:2011dq} for a comprehensive
discussion of these activities over the past two decades.

Eventually such closed strings would relax to circular
(torus-like) configurations, so-called \emph{vortons}~\cite{Davis:1988ij}. 
Typically such vortons are stabilized by coupling to (external) 
currents~\cite{Witten:1984eb,Davis:1988ip,Davis:1988jq,Davis:1995kk,
Davis:1996xs,Davis:1999ec}. The existence of vortons in a 
super--conducting scenario was supported by numerical studies some 
time ago~\cite{Lemperiere:2002en}.

It is also interesting, however, to see whether such closed strings are 
self-stabilizing in the sense that the dynamics of the underlying Lagrangian 
prevents them from contracting and eventually disappearing. 
We are particularly interested in quantum self-interactions of
a single closed string. We expect that these quantum effects become
significant for strings with radii similar to the Compton wave-lengths
of the fields within the Lagrangian. In such a scenario the string would
lose energy by emitting fluctuations of its own field rather
than by graviational radiation.

Closed strings are not subject to any non-trivial boundary condition
at spatial infinity and are therefore not stabilized topologically. On the
contrary, simple dimensional arguments show that closed strings can reduce 
their energy by shrinking. Hence they are not stabilized dynamically 
either, at least on the classical level. The na{\"\i}ve conclusion
would be that even if a closed string had existed at some earlier time
it would have decayed by emitting radiation and simultaneously
shrinking to zero size.  However, this argumentation cannot be correct at
the quantum level because narrow configurations localize the particle
in position space which, by the uncertainty principle, implies that
large momenta emerge and prevent the total energy from vanishing.

We will consider this question for the flux tube in scalar
electrodynamics and study the analog of the Nielsen-Olesen string on a
torus of radius $R$. In this case a classical instability 
occurs as $R\to0$.  The study of a toroidal string  configuration is 
significantly more complicated than that of an infinitely long  and 
axially symmetric string because the field equations cannot be simplified to  
ordinary differential equations of a single variable. This hampers the application 
of spectral methods, which have previously been used successfully to compute 
quantum effects for localized field configurations~\cite{Graham:2009zz}.
We will therefore adopt a different technique that has proven successful 
in describing similar quantum mechanical scenarios. 
It is based on the quantum mechanical uncertainty relation
\begin{equation}
\Delta p \Delta x \ge \frac{\hbar}{2}
\label{heisenberg}
\end{equation} 
and assumes that for the ground state the momentum $p$ is saturated 
by $\Delta p$ and that $\Delta x$ is a good approximation for the
spatial extension, $x$. Minimizing the Hamiltonian that is constructed 
from the replacement $p\longrightarrow \sfrac{\hbar}{(2x)}$ yields the 
correct ground state energy and extension for both the hydrogen atom (which 
is also classically unstable) and the harmonic oscillator.\footnote{For the
hydrogen atom one needs to replace $\Delta x\to2r$, where the radial coordinate 
$r$ is the distance from the center of the Coulomb potential.}

In order to apply this approach, the appropriate momentum (operator) must 
be identified. We will do this by introducing a time-dependent collective 
coordinate for the extension of the field configuration in which the 
classical instability occurs. The conjugate momentum is then extracted 
from the Lagrangian and the system is quantized by imposing the canonical 
commutation relation between the collective coordinate and its
conjugate momentum.  This so-called collective coordinate quantization
reasonably reproduces quantum properties of localized field
configurations in various examples~\cite{Weigel:2008zz}. Rather than
exploring the full quantum theory, this formulation concentrates on
the physically important modes. In the case of a collapsing cosmic string, 
the most relevant mode is obviously the radius $R$ of the torus on
which the string lives.

The paper is organized as follows. First, we complete
this introduction by describing the model of scalar electrodynamics
that we use. In section II we briefly review the Nielsen-Olesen string
in the axially  symmetric framework. In section III we provide a
detailed description of toroidal coordinates and explain the string
configuration that respects this symmetry. In particular, we derive
consistent boundary conditions. As expected from dimensional
considerations,  we find that the classical toroidal string is
unstable against shrinking to zero size.  We therefore introduce a
collective coordinate for this mode, which we quantize canonically
(section IV). We estimate the corresponding contribution to the energy
from Heisenberg's uncertainty principle and derive a quantum  energy
functional whose  minimum we seek. Since the system cannot be
decomposed into one-dimensional problems, the numerical
investigations are challenging. We relegate their description to
appendices.  Section V summarizes and concludes our studies.

\bigskip\medskip
\noindent
We consider scalar electrodynamics with the Lagrangian
\begin{equation}
\mathscr{L} = - \frac{1}{4}\,F_{\mu\nu}\,F^{\mu\nu} + \big| D_\mu\,\Phi\big|^2
- V(\Phi) \,,
\label{Lagrangian}
\end{equation}
where the covariant derivative and scalar potential are
\begin{equation}
D_\mu = \partial_\mu + i e A_\mu
\qquad {\rm and} \qquad V(\Phi) = m^2\,|\Phi|^2 +
\lambda\, |\Phi|^4\,,
\label{CoDer}
\end{equation}
respectively. To induce spontaneous symmetry breaking, we take $m^2 < 0$ and 
observe the classical vacuum expectation value (\emph{vev})
\begin{equation}
\big \langle | \Phi | \big \rangle = \sqrt{\frac{-m^2}{2\lambda}} \equiv v\,.
\label{vev}
\end{equation}
Notice that $\Phi$ and $v$ have dimension of mass in $(3+1)$ dimensions, while
$\lambda$ and $e$ are dimensionless. The (tree-level) masses of the Higgs and 
gauge fields can be identified by considering fluctuations about the $vev$ 
in the Lagrangian in eq.~(\ref{Lagrangian}),
\begin{equation} 
m_\Phi^2 = 4\lambda v^2
\,,\qquad\qquad
m_A^2 = 2 e^2 v^2\,.
\label{masses}
\end{equation}
For later reference, we briefly discuss the equation of motion for the 
temporal component of the gauge field (\emph{Gau\ss{}' law}), 
\begin{equation}
\big( - \Delta + 2 e^2 | \Phi|^2\big)\,A_0  =
i e \,(\Phi^\ast \partial_0 \Phi - \Phi \partial_0 \Phi^\ast) + 
\nabla \cdot \dot{\vek{A}}\,,
\label{gauss1}
\end{equation}
where the dot denotes a time derivative. Clearly, this equation contains no 
time derivative for ${A}_0$, and is therefore a \emph{constraint} on $A_0$ 
rather than a true dynamical equation. Furthermore, it is an elliptic differential
equation and as such has the unique solution $A_0 = 0$ (given proper boundary conditions 
at spatial infinity) if the right hand side of eq.~(\ref{gauss1}) vanishes. This 
happens if \textbf{(i)} the phase of the Higgs field is time-independent \emph{and} 
\textbf{(ii)} the gauge field $\vek{A}$ is either time-independent ($\dot{\vek{A}}=0$) 
or transverse ($\nabla\cdot\vek{A} = 0$). 
It is, of course, possible to \emph{enforce} the transversality of $\vek{A}$ (or the Weyl
condition $A_0 = 0$) by a choice of gauge, even for time-dependent fields. The respective gauge 
transformations will, however, be time-dependent and highly non-local, leading to very cumbersome 
contributions to the phase of the Higgs field and the kinetic energy of both $\vek{A}$ and $\Phi$. 
Since the kinetic energy is essential for the potential quantum stabilization of a closed string 
(cf.~section \ref{sec:QS}), it is very inconvenient to enforce the Coulomb or Weyl condition
in order to resolve Gau\ss{}' law. Instead, we have to implement eq.~(\ref{gauss1}) 
and allow for $A_0 \neq 0$ when introducing the time-dependent field configurations necessary for a quantum mechanical treatment.

\section{Infinitely long Nielsen--Olesen string}
\label{sec:NO}
Before discussing our toroidal string configuration in more detail, let us briefly 
review the well-known \emph{Nielsen--Olesen} (NO) string~\cite{Nielsen:1973cs} 
which corresponds to the limit that the (outer) torus radius becomes infinitely large.
This string is a static, infinitely long vortex 
configuration stretched along the $z$-axis, in which the Higgs field
$\Phi$ has winding number $n$, and the magnetic field points along the string axis, 
$\vek{B} \sim \vek{e}_z$:
\begin{equation}
\vek{A} = g(r)\,\frac{n}{er}\,\vek{e}_\varphi\,,\qquad\qquad
\Phi = v\, f(r)\,e^{+i n \varphi}\,.
\label{5}
\end{equation}
(We are using polar coordinates $(r,\varphi)$ in the $xy$-plane).
Notice that $\nabla\times\vek{A} = 0$ and $A_0 = 0$ for this ansatz, which is 
consistent with Gau\ss{}' law because the Higgs field $\Phi$ is time-independent.
The requirements of finite energy and continuity at the origin then
lead to boundary conditions on the profile functions $f$ and $g$,
\begin{equation}
\begin{array}{rclcl}
r \to 0 &:&  f(r) \to 0\,, &\qquad\quad& g(r) \to 0     \\[2mm]
r \to \infty &:& f(r) \to 1\,, & \qquad\quad & g(r) \to 1
\label{6}
\end{array}
\end{equation}
The NO string owes its stability to the fact that the Higgs field at spatial infinity
defines a map $S^1 \mapsto U(1)$, which in view of $\pi_1(U(1)) = \mathbbm{Z}$ cannot 
be continuously deformed into the vacuum. 

\begin{figure}[t]
\hspace*{\fill}
\includegraphics[width=7cm]{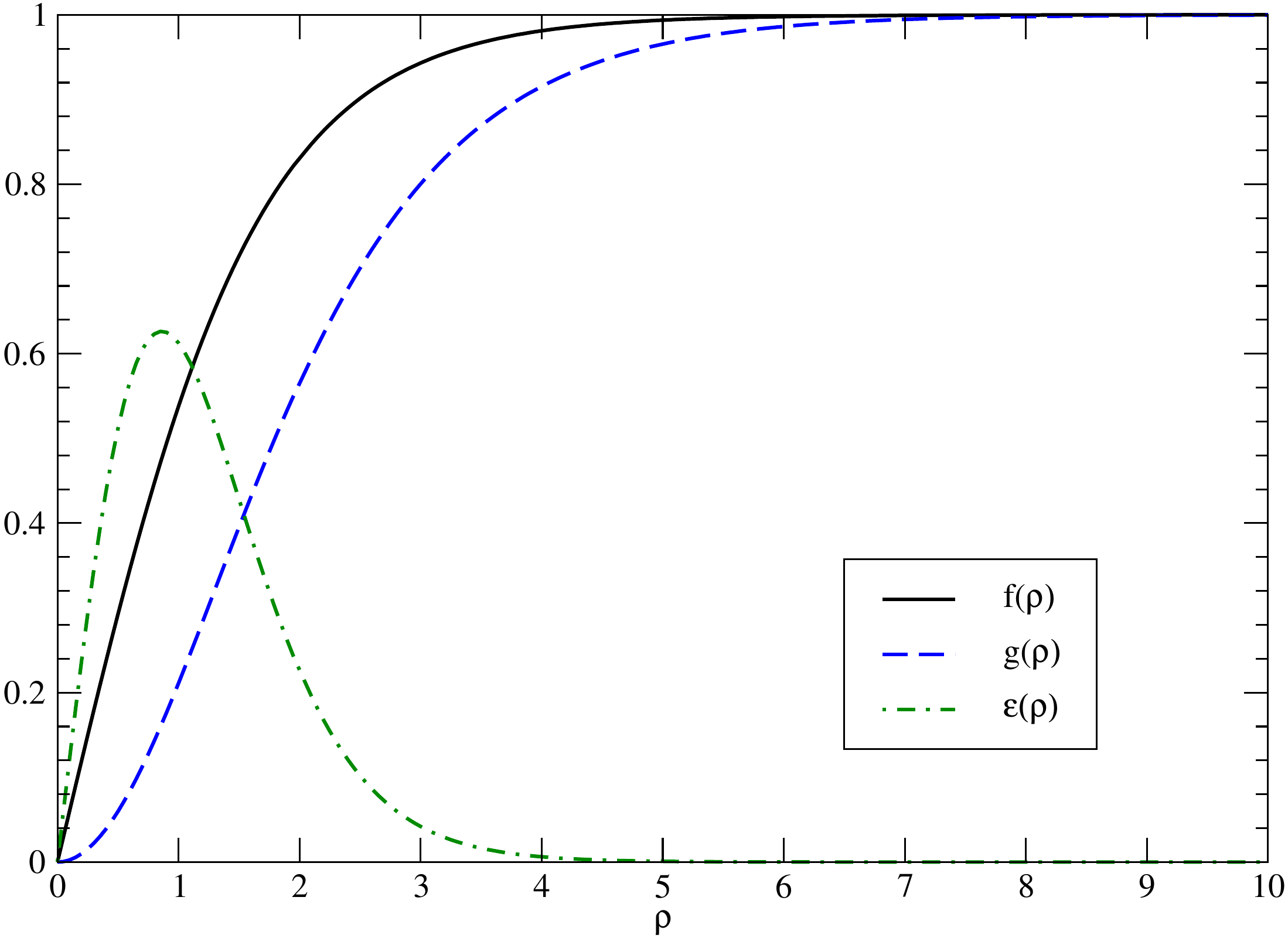}
\hspace*{\fill}
\includegraphics[width=7cm]{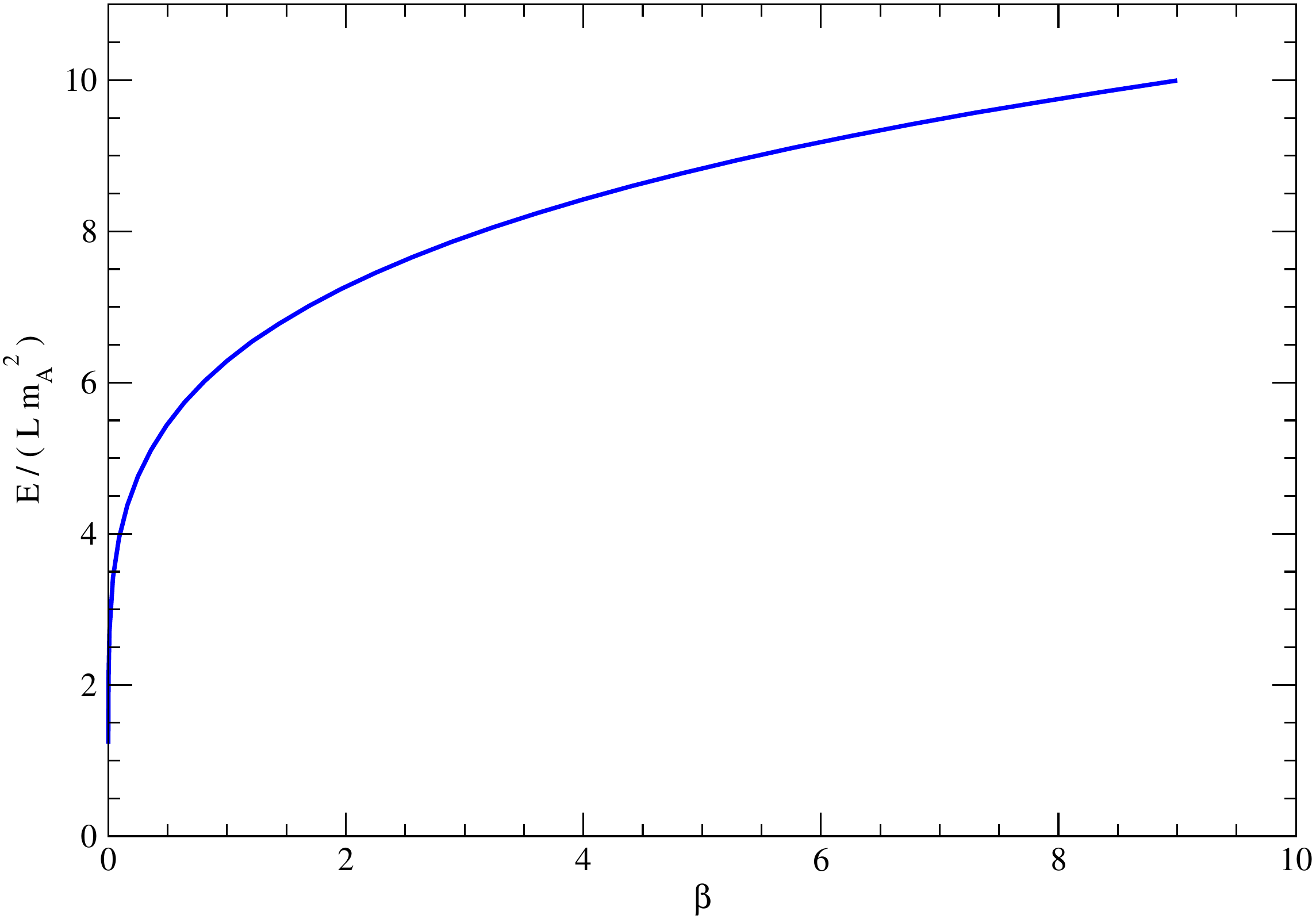}
\hspace*{\fill}
\caption{(Color online) Left panel: Nielsen--Oleson profiles for the infinitely long 
straight string for the mass ratio $\beta=1$. Also shown is the energy density, 
i.e.~the integrand of eq.~(\ref{NOeng}).
Right panel: The energy per unit length, eq.~(\ref{NOeng}), 
as a function of $\beta$ at fixed $e^2 = \frac{1}{2}$.}
\label{fig:1}
\end{figure}

We note that our model, as given by eq.~(\ref{Lagrangian}), initially has three parameters, 
the Higgs \emph{vev} $v$ and the two coupling constants $\lambda$ and $e$ (or, 
equivalently, the two masses in eq.~(\ref{masses})). After rescaling all dimensionful quantities
in units of some appropriate scale, we expect all observables to be functions of the two 
couplings $\lambda$ and $e$ independently. However, this is not always true: In the NO string, 
for instance, the profile functions only depend on the ratio
\begin{equation}
\beta = \frac{m_\Phi^2}{m_A^2} = \frac{2\lambda}{e^2} 
\label{beta}
\end{equation}
once the fundamental scale is chosen as the mass of the gauge field,
$m_A$ (and the winding number is fixed to $n=1$). This is because the gauge coupling $e$ 
scales out in an overall prefactor for the (dimensionless) energy per unit length,
\begin{align}
\frac{E}{m_A^2\,L} =
\frac{\pi}{e^2} \int\limits_0^\infty d\rho\,\rho\,\left\{ \frac{n^2}{\rho^2}\, g'(\rho)^2 
+ \Big[ f'(\rho)^2 + \frac{n^2}{\rho^2}\,f(\rho)^2\, \left(1 - g(\rho)\right)^2\Big] + 
\frac{\beta}{4}\, (1 - f(\rho)^2)^2\right \}\,,
\label{NOeng}
\end{align}
where $\rho \equiv m_A\,r$ is the scaled distance from the symmetry axis. 
The NO profiles $f(\rho)$ and $g(\rho)$ resulting from the 
minimization of this functional will only depend on the ratio $\beta$, because any change 
of couplings that leaves $\beta$ invariant can at most produce an overall prefactor to 
the functional (\ref{NOeng}), which is irrelevant for the minimization. The value of 
the (minimal) energy per unit length will then scale trivially with $e$ at fixed 
$\beta$. In figure \ref{fig:1}, we present the numerical solutions for the NO profiles 
in the case $\beta=1$, and the value of the minimal energy per unit length as a function of 
$\beta$ at fixed $e^2 = 1/2$. For our preferred values
\begin{align}
e^2 = \frac{1}{2}\,,\quad\lambda = \frac{1}{4}\qquad
\Longrightarrow\qquad m_A = m_\Phi = v\quad\mbox{and}\quad \beta = 1\,, 
\label{para1}
\end{align}
we find a minimal energy per unit length of $6.28\,m_A^2$; this will provide 
an important numerical check for the torus configuration discussed in the next section.

\section{Toroidal coordinates and strings}
\label{sec:toro}

We construct a closed string from the NO configuration by first assuming that the
latter has a finite (though very large) length $L=2\pi R$ and then identifying its two 
end surfaces. The core of the string, along which the profile functions 
$f$ and $g$ vanish, thus becomes a circle of radius $R$. Moreover, the radial 
coordinate $\rho$, {\it i.e.} the independent variable of the NO string, 
then measures the distance from this circle, and surfaces of 
constant $\rho\ne0$ are tori. We will be particularly interested in the change of 
the profile functions as the core radius $R$ decreases and becomes as small 
as the  Compton wave-lengths of the particles in the model. 
Eventually $R$ will be considered a variational parameter which will be determined
by minimizing the (quantum) energy of the torus configuration.

The NO string is characterized by a rotational symmetry about its core. This cannot be 
maintained once the string is closed, because the direction in which one moves away from 
the core does matter: If one moves away in the direction of spatial infinity, the fields 
have an infinite range to decay to their vacuum values, while there is only a finite
distance when moving towards the center of the core circle. (And it is
even not necessarily true that the fields assume their vacuum values at the center.) As a consequence, 
the lines of constant energy density must be denser on the inside of the core than on its 
outside, and since these situations are related by rotations about the core axis, 
there cannot be any axial symmetry. While the profile functions at spatial infinity 
are still subject to boundary conditions, they will result from the dynamics at the 
center of the circle. The loss of axial symmetry causes the profile functions to 
depend on more than one coordinate, which complicates matters significantly. The only 
remaining symmetry is that along the core circle.

\subsection{Geometry of the torus configuration}

The previous considerations strongly suggest the introduction of \emph{toroidal coordinates} 
which we describe in this subsection. We choose to put the circle of the string's core 
in the $xy$-plane of a Cartesian coordinate system and center it at the origin. 
The corresponding toroidal coordinates read~\cite{book_MF}:
\begin{align}
x &= R\,\frac{\sinh \tau}{\cosh \tau - \cos \sigma}\,\cos\varphi \nonumber\\[2mm]
y &= R\,\frac{\sinh \tau}{\cosh \tau - \cos \sigma}\,\sin\varphi \nonumber\\[2mm]
z &= R\,\frac{\sin \sigma}{\cosh \tau - \cos \sigma}\,.
\label{toro}
\end{align}

\begin{figure}[t]
\hspace*{\fill}
\includegraphics[width=7cm]{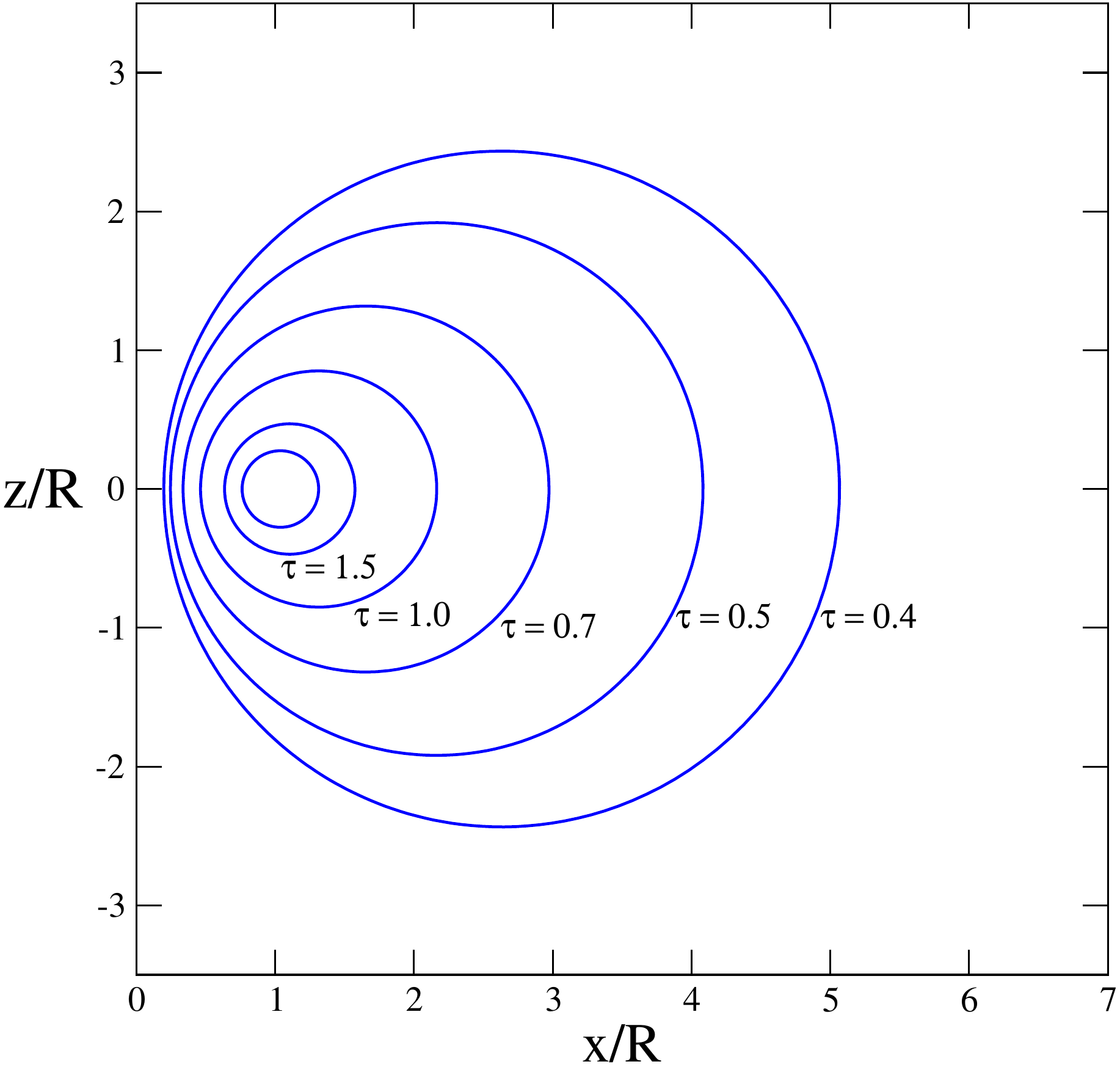}
\hspace*{\fill}
\includegraphics[width=7cm]{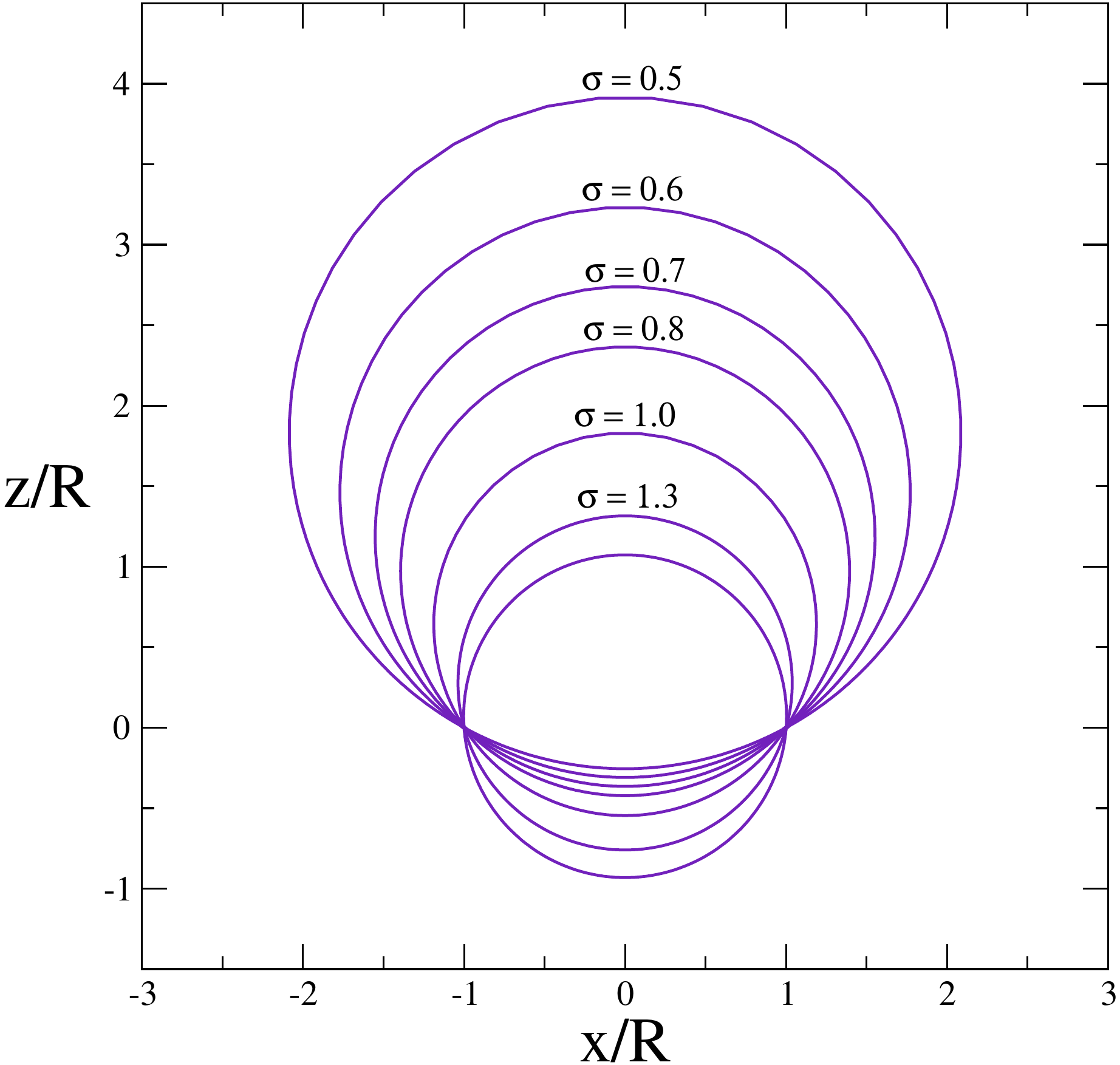}
\hspace*{\fill}
\caption{(Color online) Coordinate lines of toroidal coordinates in
the $xy$-plane ($\varphi=0$ and $\varphi=\pi$).
Left panel: Lines of constant $\tau$ (at $\varphi=0$ only, $\varphi=\pi$
is the mirror image about the $z$-axis). Right panel: Lines of constant 
$\sigma$. Each circle is composed of two distinct coordinate lines, namely 
$\sigma \in [0,\pi/2]$ (above the foci) and $(\sigma+\pi)$ (below the foci).
The remaining $\sigma$ range gives mirror images which are not displayed for 
simplicity.}
\label{fig:toro}
\end{figure}
\noindent
The coordinate $\varphi \in [0,2\pi]$ is the usual polar angle describing the  
rotation of the $xz$-plane around the vertical $z$-axis. By construction, this 
rotation is a symmetry of our configuration and $\varphi$ is a cyclic coordinate
that describes the position along the string core. The coordinate $\tau \in [0,\infty]$ 
is best understood from its iso-lines in the left panel of fig.~\ref{fig:toro}.
Smaller values of $\tau$ correspond to larger circles\footnote{These are 
indeed circles: (for $\varphi=0$) 
$[x-R{\rm coth}(\tau)]^2+z^2=R^2{\rm cosech}^2(\tau)$ is independent of $\sigma$.}
that spread out towards infinity 
on the outside, and approach the vertical axis on the inside, while the center 
of these circles gradually moves away from the torus core. The inverse $\tau^{-1}$ 
is therefore (highly non-linearly) related to the distance from the string 
core (similarly to the radius 
$\rho$ in the NO string), although it cannot be directly identified
with this quantity.\footnote{As we 
will see below, the closest analog to the radius $\rho$ in the NO string is the  
metric factor $h(\sigma,\tau)$ explained in eq.~(\ref{metric}) below.}
Finally, the angle $\sigma \in [0, 2\pi]$ describes the different points  
on the  $\tau$-coordinate lines. When viewed from the core, it plays the role
of an azimuthal angle. It is therefore natural to give the Higgs field 
winding by taking $\sigma$ as its phase.  Then the profiles are predominantly 
functions of $\tau$ only, with a $\sigma$-modulation mainly in the vicinity of the origin.
Note that origin of the coordinate system corresponds to $\tau=0$ and $\sigma=\pi$
while spatial infinity is approached as $\tau\to0$ and $\sigma\to0$.

The above considerations lead to the following ansatz for the closed string
field configuration~\cite{Postma:2007bf}
\begin{equation}
\Phi = v\,f(\sigma,\tau)\,e^{i n \sigma}\,,\qquad\qquad
\vek{A} = A(\sigma,\tau)\,\vek{e}_\sigma \,,
\label{tori:ansatz}
\end{equation}
where $\vek{e}_\sigma$ is the tangential unit vector along the lines 
in the left panel of figure \ref{fig:toro}.
To ensure finite energy, we must have $|\vek{D}\Phi|\to 0$ at large distances,
which implies 
\begin{equation}
A(\sigma,\tau) \to \frac{n}{e}\,\frac{1}{h(\sigma,\tau)} =
\frac{n}{e}\,\frac{\cosh\tau - \cos\sigma}{R}\,.
\label{tori:bc1}
\end{equation}
Here, the gradient in toroidal coordinates gives rise to the \emph{metric factor}
\begin{equation}
h(\sigma,\tau) \equiv \frac{R}{\cosh \tau - \cos\sigma}\,.
\label{metric}
\end{equation}
In view of these asymptotics, we refine our ansatz to the form
\begin{equation}
\Phi(\vek{x}) = v f(\sigma,\tau)\,e^{i n \sigma}\,,\qquad\qquad
\vek{A}(\vek{x}) = \frac{n}{e}\,\frac{g(\sigma,\tau)}{h(\sigma,\tau)}\,\vek{e}_\sigma\,,
\label{ansatz}
\end{equation}
with suitable boundary conditions on the profiles discussed below. Since the configuration 
eq.~(\ref{ansatz}) is time-independent, we can discard any temporal gauge field 
component and put $A_0=0$ to resolve Gau\ss{}' law. Note, however, that the gauge field
is no longer transverse, 
\begin{equation}
 \nabla \cdot\vek{A} = \frac{n}{e}\,\frac{1}{h(\sigma,\tau)^3}\,
\frac{\partial}{\partial\sigma}\,\Big[ g(\sigma,\tau)\,h(\sigma,\tau) \Big] \neq 0\,.
\label{transx}
\end{equation}
As expected, the magnetic field 
\begin{equation}
\vek{B}=\nabla\times\vek{A}=-\frac{n}{e}
\frac{\partial_\tau g(\sigma,\tau)}{h(\sigma,\tau)}\, \vek{e}_\varphi
\label{magfield}
\end{equation}
of the closed string points in $\vek{e}_\varphi$-direction, 
{\it i.e.} along the core of the string.

\subsection{Classical energy and boundary conditions}
\label{sec:bc}

Since our closed string ansatz is static, its energy density is just minus its 
Lagrangian density. The classical energy becomes, upon integration over space,
\begin{align}
\frac{E}{m_A} = \frac{\pi}{e^2} 
\int_0^\infty d\tau\int_{-\pi}^\pi d\sigma\,\sinh\tau\,\Bigg\{
\frac{n^2}{\RR\eta}\,&(\partial_\tau g)^2 +
\RR \eta\,\Big[ (\partial_\tau f)^2 + n^2 f^2 \,\big(1 - g\big)^2 +
(\partial_\sigma f)^2\Big] +
\nonumber \\
& + \frac{\beta}{4}\,\RR^3\,\eta^3\,\big(1 - f^2\big)^2
\Bigg\}\,.
\label{ECL}
\end{align}
As in the case of the NO string we have scaled all dimensionful quantities 
by appropriate powers of the mass of the gauge boson, $m_A$:
\begin{align}
\RR \equiv m_A\,R\,,\qquad\qquad \eta(\sigma,\tau) \equiv \frac{h(\sigma,\tau)}{R}
= \frac{1}{\cosh \tau - \cos\sigma}\,.
\label{dimparm}
\end{align}
Again, the gauge coupling $e$ only appears as an overall 
prefactor, and the profiles only depend on the mass ratio $\beta$. 
In addition we observe a non-trivial dependence on the radius $R$.

To complete the discussion of the classical torus configuration, we need to 
specify the boundary conditions for the profile functions $f$ and $g$. 
Since $\sigma$ is an angle variable and the profiles have to be single-valued, 
it is immediately clear that we must impose \underline{periodic} 
boundary conditions in $\sigma$-direction,
\begin{equation}
f(0,\tau) = f(2\pi,\tau)\,,\qquad\qquad g(0,\tau) = g(2\pi,\tau)\,.
\label{bc3}
\end{equation}
At $\tau \to \infty$, we are on the core circle where we expect the initial 
$U(1)$ symmetry to be restored, i.e.~the Higgs \emph{vev} vanishes. To avoid 
singularities at $\tau \to \infty$, the magnetic field and even the gauge potential
itself must also vanish on the core line. This enforces \underline{Dirichlet} conditions
\begin{equation}
   f(\sigma,\infty) = g(\sigma,\infty) = 0\,.
\label{bc1}
\end{equation} 
Finally, we have to find the conditions for the limit $\tau \to 0$. This corresponds 
to the largest circles in the left panel of fig.~\ref{fig:toro} and includes spatial 
infinity ($\sigma\to 0,2\pi$) and the vertical $z$-axis ($\sigma\ne 0,2\pi$). 
The first case obviously requires the vacuum configuration $f = g = 1$. Furthermore, 
when approaching the $z$-axis from any nearby point identical field configurations must be 
produced. Hence all field derivatives must vanish as $\tau\to0$. This imposes 
\underline{Neumann} boundary conditions for $\sigma\ne 0,2\pi$,
\begin{equation}
\partial_\tau f(\sigma,0) = \partial_\tau g(\sigma,0) = 0\,,
\label{bc2}
\end{equation}
Any solution of the field equations with finite total energy will automatically approach 
$f=g=1$ as $\sigma\to0,2\pi$ when $\tau=0$.  This restriction results
from the terms with positive powers of $\eta$ in eq.~(\ref{ECL}), which would diverge otherwise.

\subsection{Numerics and classical instability}
\label{sec:34}

\noindent
Application of Derrick's theorem to the covariant Lagrangian shows that no 
localized configuration is classically stable, since there is no coupling 
of negative mass dimension. We will reproduce this classical instability
explicitly by numerically calculating the minimal energy for a given
radius $R$. This will provide us with important information and techniques
needed to subsequently investigate quantum-mechanical effects.

To minimize the energy in eq.~(\ref{ECL}) subject to the boundary conditions,
eqs.~(\ref{bc3})--(\ref{bc2}), 
we discretize the coordinate space by a square lattice of $(N_s+1) \times (N_t+1)$ grid points,
\begin{align}
\tau &= \tau_0 + t\,\frac{\tau_\infty-\tau_0}{N_t}\,, & t &= 0,\ldots,N_t 
\nonumber\\[2mm]
\sigma &= \displaystyle s\,\frac{2 \pi}{N_s}\,,  &s &= 0,\ldots,N_s \,.
\label{lat1}
\end{align}
For our numerical treatment $\tau=0$ and $\tau=\infty$ have been replaced 
by finite values $0\lesssim\tau_0 \ll 1$ and $\tau_\infty \gg 1$, respectively.
The discretization replaces the continuous profiles by indexed quantities
\begin{align}
f_{st} \equiv f(s\,\Delta_s, \tau_0 + t\,\Delta_t)\,,\qquad\qquad
g_{st} \equiv g(s\,\Delta_s, \tau_0 + t\,\Delta_t)
\label{profdis}
\end{align}
defined on the lattice sites. The grid spacings are
$\Delta_t \equiv (\tau_\infty-\tau_0) / N_t$ and $\Delta_s \equiv 2 \pi / N_s$,
respectively, which can be made arbitrarily small by increasing the
lattice size. With the integrals replaced by Riemann sums, the minimization 
of the classical energy (\ref{ECL}) turns into a high-dimensional discrete 
minimization problem for the variables $f_{st}$ and $g_{st}$, which we solve by 
a combination of standard algorithms such as iterated overrelaxation and simulated 
annealing. Further details on the discretization and the minimization algorithms 
can be found in appendix \ref{app:lattice}.

\medskip
\noindent
Our model has a number of parameters, which can be partitioned into two classes:
\begin{enumerate} 
\item \emph{physical parameters}: These are the gauge coupling $e$, the 
ratio of couplings $\beta$ given in eq.~(\ref{beta}), and the
dimensionless core radius $\RR$. As explained earlier, the gauge
coupling only appears as an overall prefactor in the energy once
dimensionless variables have been introduced.
We will therefore adopt $e^2=1/2$ as a standard choice and study the energy 
as a function of $\beta$ and $\RR$.
\item \emph{numerical parameters}:  These are related to the discretization and 
should become immaterial in the continuum limit $N_s\to\infty$ and $N_t\to\infty$. 
In addition to the grid size the numerical treatment 
also contains the cutoffs $\tau_0$ and $\tau_\infty$ for the 
$\tau$-parameter. In practice, the cutoffs are replaced by
\begin{align}
d_{\rm min} =  \frac{2}{e^{\tau_\infty}+1}\,,\qquad\qquad
d_{\rm max} =  \frac{2}{e^{\tau_0}-1}\,,  
\label{dminmax}
\end{align}
where $Rd_{\rm min/max}$ are the distances of the $\tau$-isoline from the torus core, 
measured on the $x$-axis inside ($d_{\rm min}$) or outside ($d_{\rm max}$), 
cf.~fig.~\ref{fig:toro}. Hence the continuum limit also requires 
$d_{\rm min}\to0$ and $d_{\rm max}\to\infty$.
\end{enumerate}

First, we verify that the numerical (discretization) parameters become irrelevant 
in the continuum limit. This exercise also suggests appropriate data for 
the numerical parameters to be used in later computations.
To this end, we hold all of them fixed at reasonable values
\begin{align}
N_s = N_t = 300\,,\qquad\qquad
d_{\rm min} = 10^{-3}\,,\qquad\qquad
d_{\rm max} = 20.0
\label{reasonable} 
\end{align}
and vary one at the time. The physical parameters are fixed to $\RR = 10$ and
$\beta = 1$ in all cases, so that distances are all measured in units of the 
Compton wavelength of both the Higgs or the gauge boson (which are equal). 

\begin{table}[ht]
\centering
\begin{tabular}{|c||c|} \hline
$d_{\rm min}$ \rule{0mm}{5mm} & $\EE = E/m_A$ \nonumber 
\\  \hline\hline
$10^{-1}$ &  394.09 \\ \hline
$10^{-2}$ &  390.75 \\ \hline
$10^{-3}$ &  390.09 \\ \hline
$10^{-4}$ &  389.47 \\ \hline
$10^{-5}$ &  388.85 \\ \hline
$10^{-6}$ &  388.23 \\ \hline
$10^{-7}$ &  387.62 \\ \hline
$10^{-8}$ &  387.01 \\ \hline
$10^{-10}$ &  --- \\ \hline
\end{tabular}
\hspace*{1cm}
\begin{tabular}{|c||c|} \hline
$d_{\rm max}$ \rule{0mm}{5mm} & $\EE = E/m_A$ \nonumber 
\\  \hline\hline
12 & 388.08  \\ \hline
14 & 388.05  \\ \hline
16 & 388.02  \\ \hline
18 & 388.00  \\ \hline
20 & 387.98  \\ \hline
25 & 387.94  \\ \hline
30 & 387.90  \\ \hline
40 & 387.86  \\ \hline
50 & 387.83  \\ \hline 
\end{tabular}
\hspace*{1cm}
\begin{tabular}{|c||c|} \hline
$N_s = N_t$ \rule{0mm}{5mm} & $\EE = E/m_A$ \nonumber 
\\  \hline\hline
100 & 383.35  \\ \hline
120 & 384.89  \\ \hline
150 & 386.43  \\ \hline
180 & 387.46  \\ \hline
200 & 387.98  \\ \hline
250 & 388.98  \\ \hline
300 & 389.52  \\ \hline
500 & 390.28  \\ \hline
800 & ---     \\ \hline 
\end{tabular}
\caption{Energy dependence on numerical parameters. A dash indicates that 
the relaxation did not converge in the predefined number of steps ($50000$).}
\label{tab:1}
\end{table}

\medskip
Tab.~\ref{tab:1} shows the results. The largest dependence is on the short distance 
cutoff $d_{\rm min}$ and the lattice size $N_s = N_t$, but only if these are taken 
too far from the continuum limit. For conservative values $d_{\rm min} \le 0.01$ 
and $N_s = N_t \ge 150$, the result for the total energy changes
at most by $1.1 \%$ when varying any of the numerical parameters. Combined with the 
systematic error from the relaxation algorithm (which may not find the 
\emph{absolute} minimum), we estimate an overall accuracy of about $2 \%$ 
for our classical energy calculations. (This estimate is not indicated by error bars 
in the plots below.) The conclusion is that the numerical parameters in
eq.~(\ref{reasonable}) are sufficient to reach the continuum limit, at least 
as long as the particle masses and $\RR$ are not chosen to have
extreme values. 

{The right panel of figure~\ref{fig:1a} shows that the straight line (or NO) limit is 
reached for radii only slightly larger than the Compton wave-length of the gauge field.
This suggests that energy loss from radiating gauge and/or Higgs fluctuations
is indeed irrelevant for simulations of networks of closed strings that refer to
cosmological length scales.

\begin{figure}[t]
\includegraphics[width = 7.5cm]{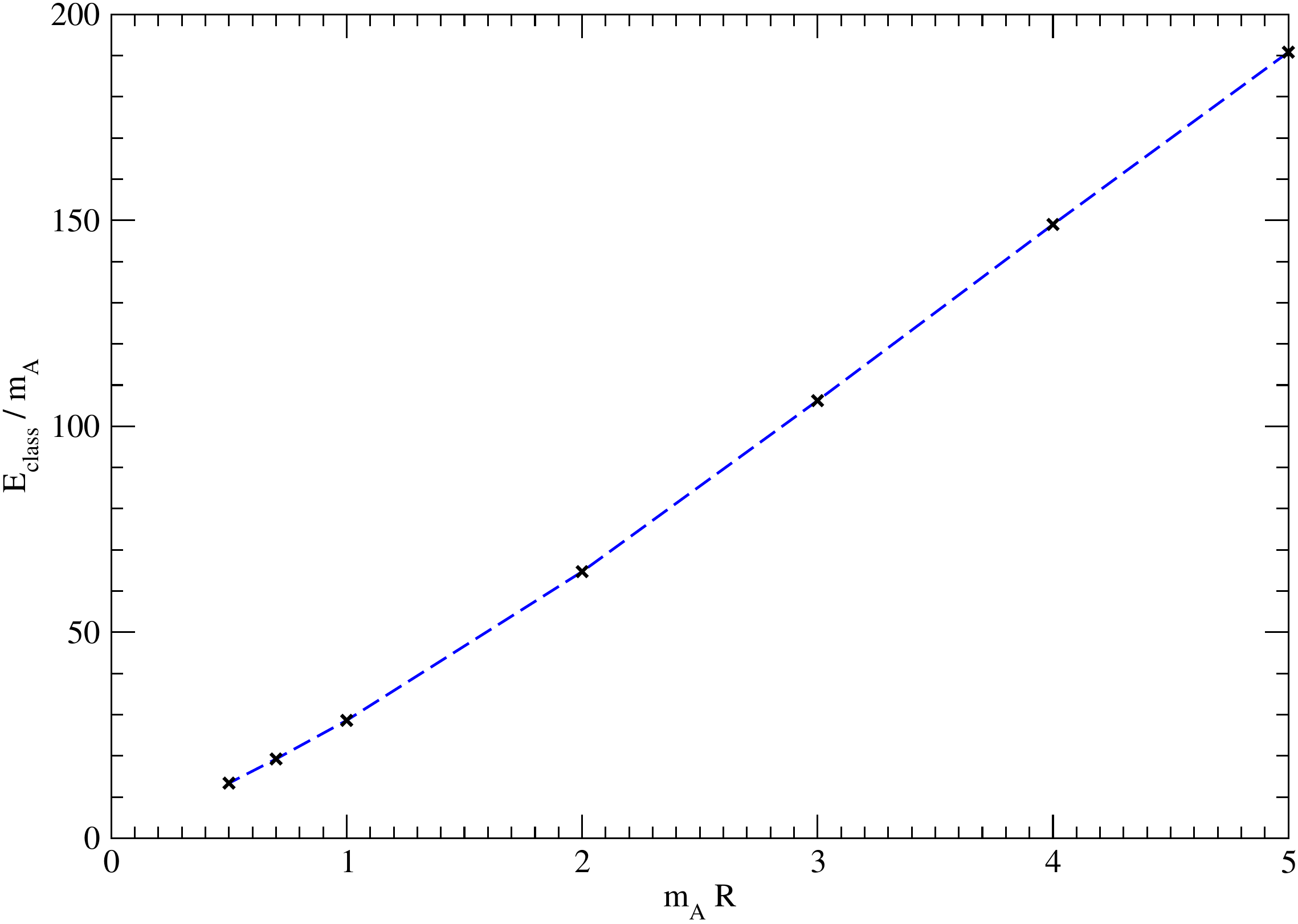} \hspace*{\fill}
\includegraphics[width = 7.5cm]{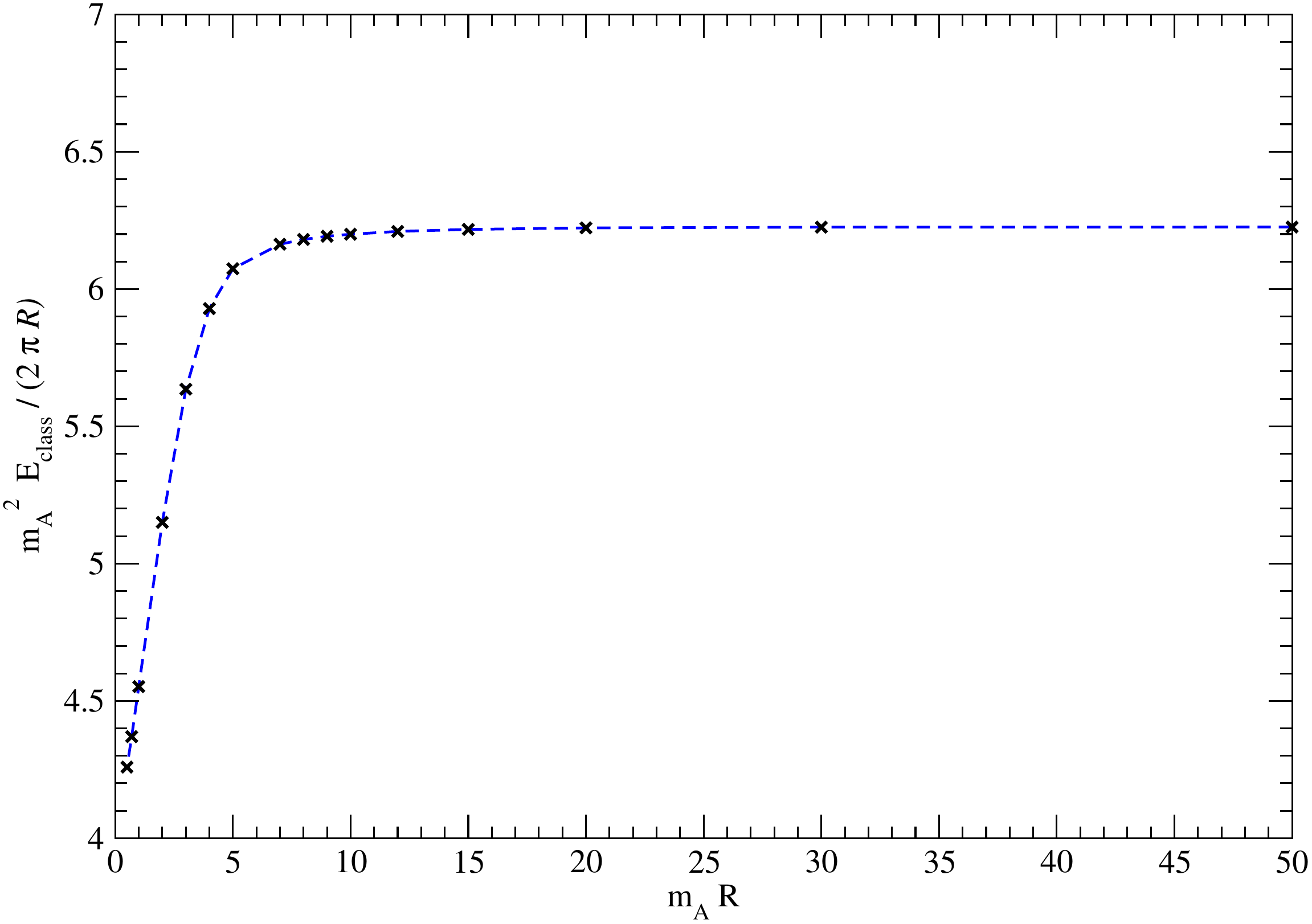}
\caption{(Color online) Dependence of the total energy (left panel) and the 
energy per unit length (right panel) on the torus radius, 
at otherwise fixed parameters.}
 \label{fig:1a}
\end{figure}

Next, we want to demonstrate that the torus configuration has a classical instability 
against shrinking into nothing ($R \to 0$). To see this, we leave the discretization 
parameters at their values in eq.~(\ref{reasonable})
and fix $\beta=1$ (and $e^2 = \frac{1}{2}$).\footnote{For very small radii, 
when opposite sides of the torus start to interact, we need a better resolution in the area 
inside the torus and near the torus core. To do so, we must enlarge the effective $\tau$-window 
and thus $[d_{\rm min},d_{\rm max}]$ while also increasing the 
lattice extensions to keep the discretization errors small. We have varied the 
numerical parameters in such cases until a stable result was found.}
In fig.~\ref{fig:1a}, we see that the total energy is a linear function of 
the torus radius $R$ down to very small radii, of the order of a few Compton
wave lengths. A better signal is obtained by studying the total energy per unit (core) length, 
$E / (2 \pi R)$. As can be seen from the right panel in fig.~\ref{fig:1}, the energy per 
unit length saturates at a constant asymptotic value (which equals the energy per unit
length of the NO string listed after eq.~(\ref{para1})), down to about $\RR \lesssim 5$. 
For smaller radii, the interaction between opposite sides of the torus leads to a 
considerable drop in the energy per unit length (cf.~density plots below). 
As a consequence, the total energy $E$ of the torus configuration decays more than 
linearly at small core radii $R \to 0$, which means that it is classically unstable.

\medskip\noindent
We can also verify that the profiles are predominantly functions of 
$\tau$, with only a very small $\sigma$-modulation. This is shown in the left 
panel of figure \ref{fig:3} for the Higgs profile $f(\sigma,\tau)$ (the
gauge boson behaves similarly). A notable variation with $\sigma$ is only visible at 
very small $\tau$ when $\sigma \approx 0$ (i.e. on the vertical axis near the origin).
The right panel of fig.~\ref{fig:3} demonstrates that the NO profiles obtained in the
last section agree with the torus profiles in the limit $\RR \gg 1$, provided that we 
identify the radius $r$ of the NO configuration, eq.~(\ref{5}), 
with the outer distance $2R/({\rm e}^\tau-1)$, cf. eq.~(\ref{dminmax}).

\begin{figure}[!t]
\includegraphics[width = 7.5cm, height=6.5cm]{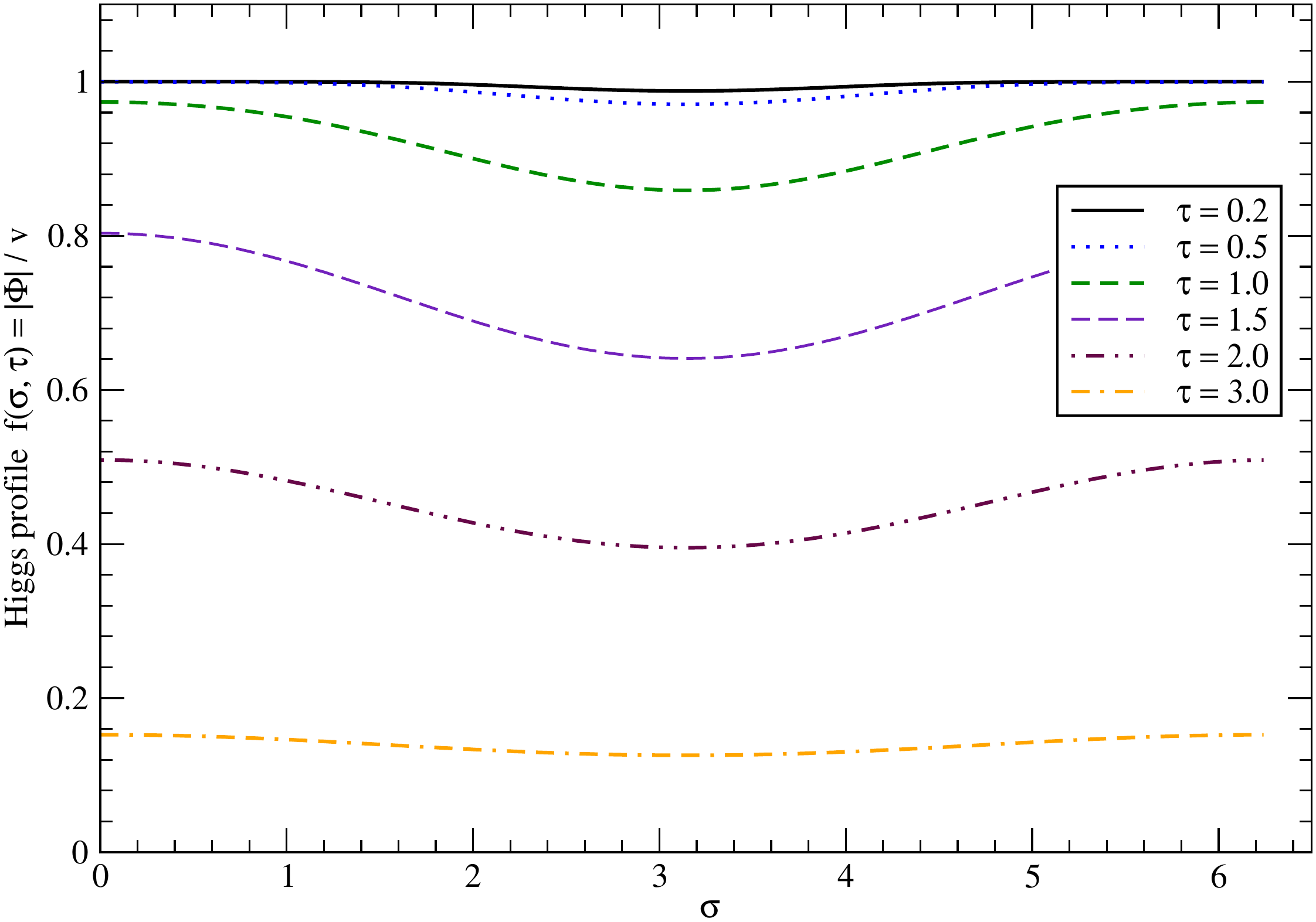} \hspace*{\fill}
\includegraphics[width = 7.5cm, height=6.5cm]{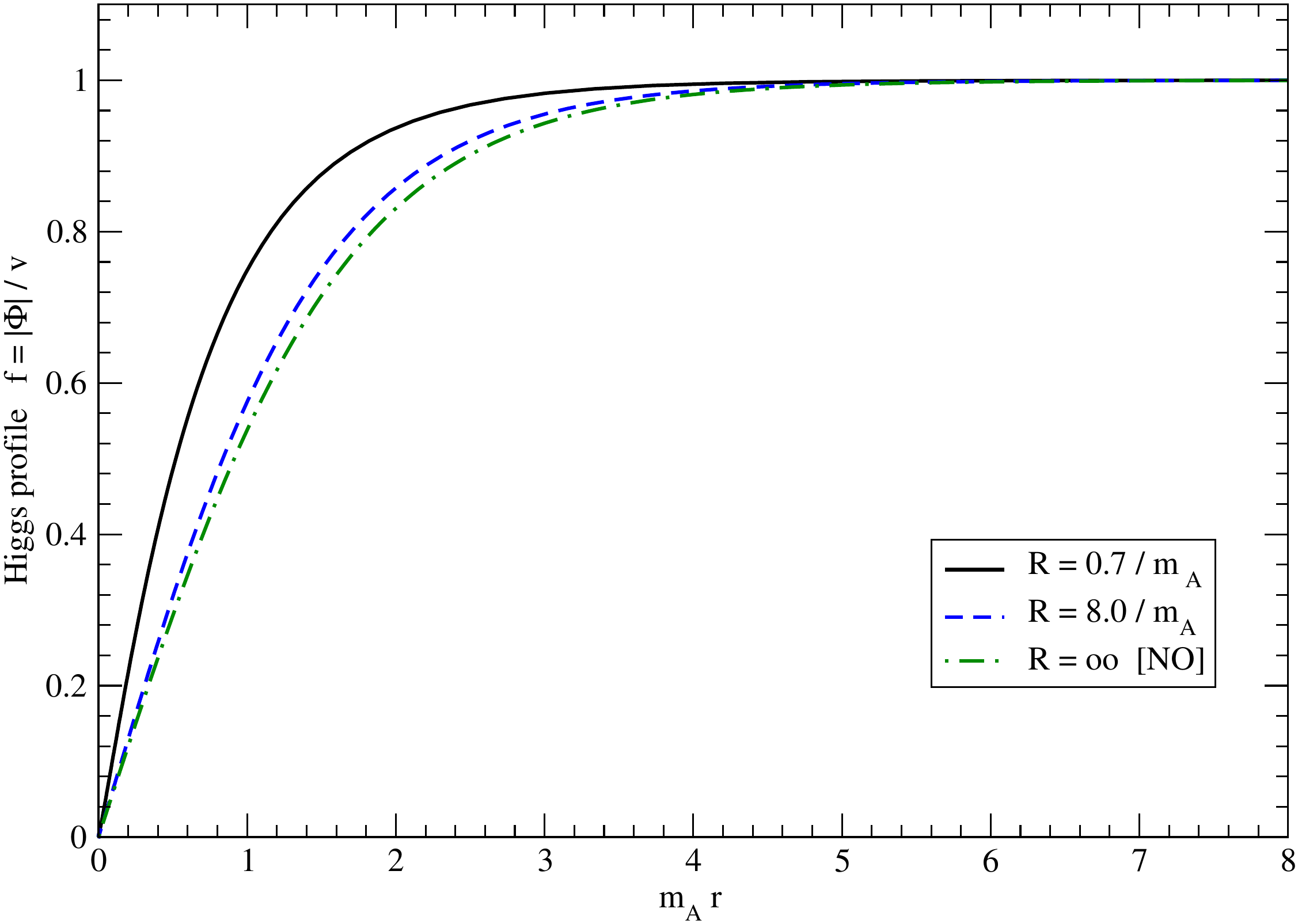}
\caption{(Color online) Left panel: Higgs profile $f(\sigma,\tau)$ as a function of 
$\sigma$ for $\hat{R} = 8.0$ and various values of $\tau$.  Right panel: Comparison 
of the NO Higgs profile from section \ref{sec:NO} with the profile from the 
torus configuration for two values of $\RR$. Both charts refer to the model 
parameters $e^2 = \frac{1}{2}$ and $\beta = 1$. In the right panel the variable 
$r$ refers to $r$ as in eq.~(\ref{5}) for the NO string and to $2R/({\rm e}^\tau-1)$
for the closed string.}
\label{fig:3}
\end{figure}

\medskip\noindent
Finally, we visualize our configuration with density plots of the Higgs magnitude 
$|\Phi|/m_A \sim f(\sigma,\tau)$ in a plane perpendicular  to the torus 
core.\footnote{Here and in the following, the toroidal coordinates
are translated back to Cartesian distances for better visualization.}
(The real configuration is found by rotating the 2D density plot around 
the vertical axis.) From the Higgs magnitude, we see the expected profiles 
for large radii, and the overlap in the inner region of the torus for very small 
radii. The other local quantities (magnetic field, energy density etc.) show a 
qualitatively similar behavior.

\section{Quantum stabilization}
\label{sec:QS}

We have seen above that the energy of the torus decreases linearly for a 
wide range of torus core radii $R$ (and even faster than linearly) at very 
small $R$), {\it i.e.} there is a classical instability of the torus configuration 
against shrinking into nothing. As argued from the uncertainty principle in the
introduction, we expect that quantum effects will eventually stabilize the 
torus at a sufficiently small radius, so that a stable quantum torus 
configuration emerges.

\subsection{Quantum action for the instable mode}
\noindent
There are various ways to compute the relevant quantum corrections. 
In situations where the number of external degrees of freedom or 
quantum numbers is large, the leading corrections come from the 
small amplitude fluctuations, which can be efficiently computed using
\emph{spectral methods}~\cite{Graham:2009zz}, at least when
there is sufficient symmetry to formulate a scattering problem. 
Formally, this correction is of order $\mathscr{O}(\hbar)$. In the present 
case, however, the instability occurs in a very specific channel,
associated with the \emph{scale} of the configuration. This strongly suggests that 
collective quantum fluctuations in this channel may be significant and eventually
prevent the configuration from collapsing. To identify these quantum 
fluctuations, we promote the torus core radius $R$ to a dynamical variable
by making it time-dependent, $R=R(t)$.

More precisely, we consider a time-independent field configuration 
$\Field_R(\vek{x})$ (not necessarily the one that minimizes  eq.~(\ref{ECL})), 
where $\Field$ stands for any of the fields in eq.~(\ref{ECL}). The 
subscript $R$ indicates that the configuration may depend implicitly
(through its functional form) on the torus radius $R$.
It is important to note that this dependence does \emph{not} arise from the use of 
toroidal coordinates, which is merely a parameterization of the space 
point $\vek{x}$, but rather describes a parameter in the functional form of 
$\Field_R$ which is independent of the coordinates used.
If we now promote $R$ to a dynamical variable by scaling 
$R \to \lambda(t) R \equiv R(t)$, we obtain a time-dependent field configuration
\begin{align}
 \Field(\vek{x},t) = \Field_{R(t)}(\vek{x})\,,
\end{align}
where $\vek{x}$ is time-independent.
The general form of the action associated with the collective coordinate $R(t)$ must be
\begin{equation}
S = \int d^4x \,\mathscr{L}[\Field_{R(t)}]
= \int dt\,\left[ \frac{1}{2}\,u(R)\,\dot{R}^2 - E(R) \right]\,.
\label{504}
\end{equation}
\noindent
Here, the dot indicates a time derivative and we have explicitly indicated that 
the functions $u$ and $E$ depend on the collective coordinate $R(t)$. In addition they 
also depend on the model parameters\footnote{In units of the gauge mass, 
$\uu = u/m_A$ and $\EE = E/m_A$ both depend on the gauge coupling 
only by an overall prefactor of $1/e^2$. As explained earlier, this means that the 
quantum energy scales trivially with $e$ and the optimal profiles will depend on the 
coupling constants only through the mass ratio $\beta$, cf.~eq.~(\ref{beta}).} 
and the choice of profile functions. For a time-independent radius,
the action reproduces the energy functional in
eq.~(\ref{ECL}). Moreover, the time derivative $\dot{R}$ appears quadratically because of time-reversal invariance 
and because the Lagrangian has at most two field derivatives. 

\begin{figure}[t]
\includegraphics[width=8cm,height=7.5cm]{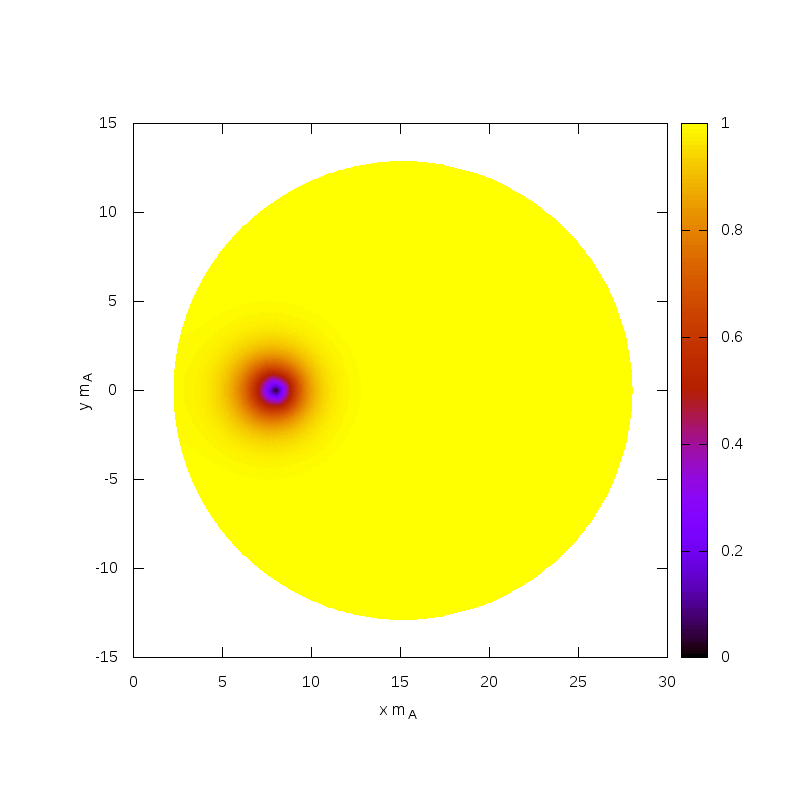} \hspace*{\fill}
\includegraphics[width=8cm,height=7.5cm]{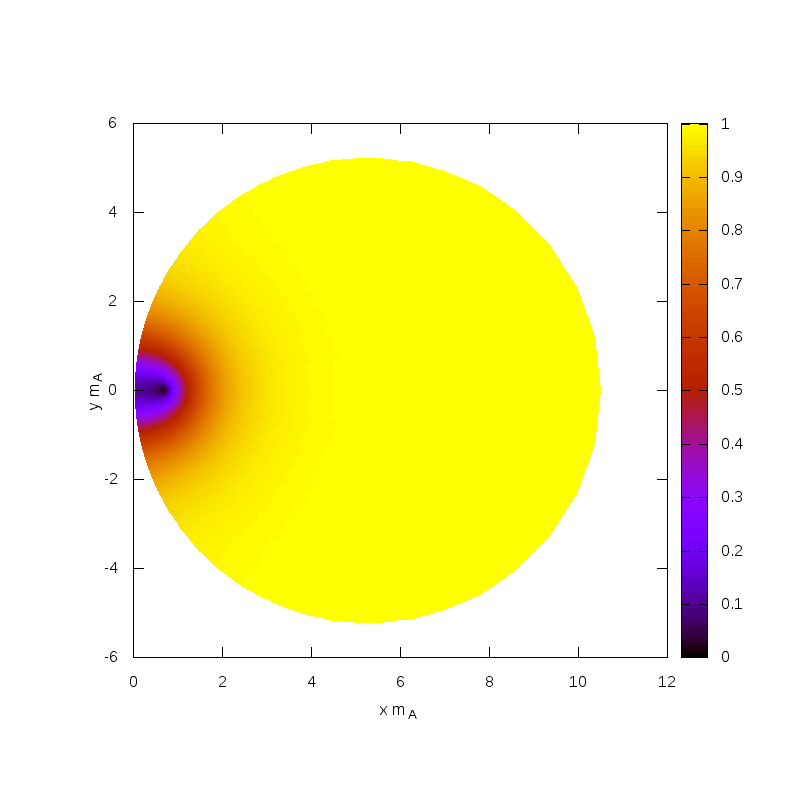}
\caption{(Color online) Density plots of the Higgs magnitude in a plane perpendicular to 
the torus. All positions are in units of the gauge boson Compton wavelength $m_A^{-1}$.
In the left panel, the torus has a radius $\RR = m_A R = 8$, while in
the right panel shows results for $\RR = 0.7$. Notice the different scales 
on the axes in both cases.}
\label{fig:new}
\end{figure}

The profile functions depend on the spatial coordinates only via the ratio $\vek{x}/R$ 
that defines the toroidal coordinates, {\it cf.} eq.~(\ref{toro}). We elevate this radius 
to a time dependent variable. This time dependence produces non-zero time derivatives
in the action, eq.~(\ref{504}), from which we read off the mass function $u(R)$.
In addition to the scale dependence of the profile functions the fields may be 
subject to explicit scaling as well. 

The vacuum expectation value of the Higgs fields is fixed at spatial infinity. Thus
there is no explicit scaling in this case and the parameterization of the time dependent
Higgs field becomes,
\begin{equation}
\phi(\vek{x},t)=\phi_0(\vek{x}/R(t))=
\widetilde{\phi}_0(\sigma_t,\tau_t)\,.
\label{scalH1}
\end{equation}
While $\widetilde{\phi}_0(\sigma,\tau)$ is a prescribed profile function of toroidal 
coordinates, $\sigma_t$ and $\tau_t$ denote the particular toroidal coordinates for 
the scaled point $\vek{x}/R(t)$. Since we consider the field $\phi(\vek{x},t)$ at a 
given space point $\vek{x}$, we must treat the arguments of the profile function
$\widetilde{\phi}_0$ as time-dependent variables. 
As we anticipated, this dependence induces a non-zero time derivative
that is most conveniently computed from eq.~(\ref{toro}) for $\varphi=0$, taking into
account that $\partial \vek{x}/\partial t=0$:
\begin{equation}
\begin{pmatrix}\dot{\sigma}_t \cr \dot{\tau}_t\end{pmatrix}
=-\frac{\dot{R}}{R}\left[\begin{pmatrix}
\frac{\partial x}{\partial \sigma} &  \frac{\partial x}{\partial \tau}\\[2mm]
\frac{\partial z}{\partial \sigma} &  \frac{\partial z}{\partial \tau}
\end{pmatrix}^{-1} \begin{pmatrix} x \cr z \end{pmatrix} 
\right]_t
=\frac{\dot{R}}{R}\begin{pmatrix}
\sin\sigma_t\,\cosh\tau_t \cr
\cos\sigma_t\,\sinh\tau_t \end{pmatrix}\,.
\label{chainrule}
\end{equation}
The subscript on the square bracket indicates that the toroidal 
coordinates are to be taken at the time dependent values
$\sigma=\sigma_t$ and $\tau=\tau_t$.
Then the time derivative of the Higgs fields becomes
\begin{equation}
\dot{\phi}(\vek{x},t)=
\frac{\dot{R}}{R}\left[\deriv \widetilde{\phi}_0(\sigma,\tau)
\right]_t
\label{scalH2}
\end{equation}
where we have defined the dimensionless radial derivative
\begin{equation}
\deriv=\sin\sigma\,\cosh\tau\,\frac{\partial}{\partial \sigma} +
\cos\sigma\,\sinh\tau\,\frac{\partial}{\partial\tau}\,,
\label{deriv}
\end{equation}
which does not depend on the radius $R$.
For the gauge field the situation is a bit more complicated because
there is no boundary condition that disallows an overall scaling by
an $R$ dependent function. In principle any parameterization would
be allowed, because as a solution to the full field equations the
correct $R$ dependence would be a result. However, we consider
only a subset of these equations, specified by our ansatz. As for the 
NO string the field equations require that the covariant derivative 
vanishes as spatial infinity. On the other hand we do not want to abandon 
our choice $g(\sigma,\tau)\to1$ for the boundary condition at 
$|\vek{x}|\to\infty$. Since the derivative induces a factor $1/R$ 
due to $\frac{\partial}{\partial\vek{x}}
=\frac{1}{R}\frac{\partial}{\partial(\vek{x}/R)}$ we include 
exactly this factor in the parameterization of the gauge field 
\begin{equation}
\vek{A}(\vek{x},t)=\frac{1}{R(t)}\,\widetilde{\vek{A}}_0
(\sigma_t,\tau_t)
\label{scalA1}
\end{equation}
to ensure that the covariant derivative has a uniform scale dependence. This 
factor additionally contributes to the time derivative of the gauge field
\begin{equation}
\dot{\vek{A}}(\vek{x},t)=\frac{\dot{R}}{R^2}
\left[(\deriv -1)\,\widetilde{\vek{A}}_0(\sigma,\tau)
\right]_t\,.
\label{scalA2}
\end{equation}
To obtain the Lagrange function, eq.~(\ref{504}), we need to integrate 
over coordinate space with the measure
\begin{equation}
d^3x = R^3 d^3[x/R] = R^3\, d^3[\sigma_t,\tau_t,\varphi_t]
\label{scalint}
\end{equation}
where the last expression refers to the integration measure of toroidal 
coordinates. After spatial integration the scale dependence factors as 
powers $R^k$ and the ratio $\dot{R}/R$. In summary, the ans\"atze for 
the field configuration together with their time derivatives are
\begin{equation}
\begin{array}{llll}
\Phi(\vek{x},t)&=v\,{\rm e}^{i n \sigma} f(\sigma,\tau)\,,
\qquad
&\dot{\Phi}(\vek{x},t) &=v\, {\rm e}^{i n \sigma}\,\frac{\dot{R}}{R}\,
\Big[i n f(\sigma,\tau)\,\sin\sigma\,\cosh\tau+\deriv f(\sigma,\tau)\Big]\,,
\\[4mm]
\vek{A}(\vek{x},t)&=\frac{n}{e}\,
\frac{{\scriptstyle g}(\sigma,\tau)}
{R\,\eta(\sigma,\tau)}\,\vek{e}_\sigma\,,
&\dot{\vek{A}}(\vek{x},t) &=\frac{n}{e}\,\frac{\dot{R}}{R}\,
\Big[\deriv - 1\Big]\,
\frac{{\scriptstyle g}(\sigma,\tau)}
{R\, \eta(\sigma,\tau)}\,\vek{e}_\sigma \,.
\end{array}
\label{tderiv}
\end{equation}
We have omitted the index on the toroidal coordinates because we have turned 
them into (dummy) integration variables, as in eq.~(\ref{scalint}). 
Also, we have not explicitly indicated the 
implicit dependence of the profiles on the radius induced by the stationary
conditions. Note that to facilitate the numerical calculation, the 
scaling has been set up such that these profile functions obey exactly
the boundary  conditions discussed in section~\ref{sec:bc}.

The kinetic term $\dot{\vek{A}}^2$ for the gauge boson thus becomes
\begin{align}
\dot{\vek{A}}^2 = \left(\frac{n}{e}\right)^2\,\frac{\dot{R}^2}{2 R^4\,\eta^2}
\,\Big[ \big( \cos(2\sigma) + \cosh(2 \tau) \big)\,g^2 + 
4 \,\cos\sigma\,\cosh\tau g \deriv g  
+ 2 \,\big(\deriv g\big)^2 \Big] 
\label{scalx}
\end{align}

\subsection{Gau\ss{}' law}
\noindent
For time-dependent fields, the initial Lagrangian eq.~(\ref{Lagrangian}) can be split
into three pieces:
\begin{align}
\mathscr{L} &= \Big \{ \frac{1}{2}\,\dot{\vek{A}}^2 + | \dot{\Phi} |^2 \Big \} 
+ \Big \{ \frac{1}{2}\,(\nabla A_0)^2 + e^2 \,|\Phi|^2\,A_0^2 - A_0\,\mathcal{J}_0 \Big \}
- \Big \{ | \Vek{D} \Phi |^2 + V(\Phi) \Big \} \nonumber\\[2mm]
&= \mathscr{L}_{\rm kin} + \mathscr{L}_{\rm gauss} + \mathscr{L}_{\rm stat}\,. 
\label{lagrangex}
\end{align}
The $A_0$ field only appears in $\mathscr{L}_{\rm gauss}$, and this contribution gives 
rise to Gau\ss{}' law, eq.~(\ref{gauss1}), with the source
\begin{align}
\mathcal{J}_0\equiv i e \,\Big( \dot{\Phi}\,\Phi^\ast - \dot{\Phi}^\ast\,\Phi\Big) + 
\nabla \dot{\vek{A}}\,.
\end{align}
In the static case, we had $\mathcal{J}_0 = 0$ and hence $A_0 = 0$ and $\mathscr{L}_{\rm gauss} = 0$.
This is no longer the case when the fields are time-dependent because of the dynamic 
scaling of the closed string. From $\dot{\Phi}$ in eq.~(\ref{tderiv}), 
the Higgs contribution to the source $\mathcal{J}_0$ reads (recall $m_A=\sqrt{2}ev$)
\begin{align}
i e \,\Big( \dot{\Phi}\,\Phi^\ast - \dot{\Phi}^\ast\,\Phi\Big) = 
- \frac{n}{e}\,m_A^2\,\frac{\dot{R}}{R}\,f^2\,\sin\sigma\,\cosh\tau\,.
\label{charge1} 
\end{align}
The gauge field contribution to $\mathcal{J}_0$ is a bit more involved. 
Since our torus ansatz in eq.~(\ref{ansatz}) is not transverse, we can 
either take the time derivative of eq.~(\ref{transx}) following the rule (\ref{tderiv}),
or take the divergence of $\dot{\vek{A}}$ in eq.~(\ref{tderiv}). 
Using the identity $[\nabla,\deriv]=-\nabla$, the result from both 
calculations is
\begin{align}
\nabla \cdot \dot{\vek{A}} = \frac{n}{e}\,\frac{\dot{R}}{R^3}\,\big(\deriv - 2\big)\,
\Big[ \eta^{-3}\,\partial_\sigma(g\,\eta)\Big]\,. 
\label{charge2}
\end{align}
From eqs.~(\ref{charge1}) and (\ref{charge2}), we observe $\mathcal{J}_0 \sim \dot{R}/e$ 
and hence $e\,A_0 \sim \dot{R}$ due to the field equation (\ref{gauss5}).
The time dependent scaling thus
induces a temporal component of the gauge field that is proportional to 
the time derivative of the radius. It is thus suggestive to set
\begin{equation}
e\,A_0(\vek{x},t)= \frac{\dot{R}}{R}\,q(\sigma,\tau)\,,
\label{a0}
\end{equation}
with a dimensionless profile function $q$ that solves, 
in principle,
\begin{align}
\Big[ (-\Delta) + m_A^2 f^2 \Big]\,q = \mathcal{J}_0 e\,\frac{R}{\dot{R}}
\equiv \mathcal{Q}[f,g] \,.
\label{gauss5}
\end{align}
Instead of solving eq.~(\ref{gauss5}) directly, it is much more 
efficient to treat Gau\ss{}' law on equal footing with the remaining
field equations.  That is, we
keep $q$ as an independent field that appears in the contribution
\begin{align}
\int d^4x \,\mathscr{L}_q = \int d^4 x\, \frac{\dot{R}^2}{e^2\,R^2}\,\left[ \frac{1}{2}
\left( (\nabla q)^2 + m_A^2\,f^2\,q^2\right) - q \mathcal{Q}\right] \equiv 
\int dt\,\frac{1}{2}\,u_q(R)\,\dot{R}^2 
\label{supergauss}
\end{align}
to the kinetic energy in eq.~(\ref{504}). Since the remaining parts of the action 
are independent of $q$, the minimization of the action with respect
to $q$ will minimize the kinetic term 
and thus $u_q$, which is equivalent to Gau\ss{}' law eq.~(\ref{gauss5}). The conclusion is 
that treating $f$, $g$, and $q$ as independent fields will automatically give the correct 
contribution to the (quantum) energy from Gau\ss{}' law, provided that $q$ is determined 
from the minimization of $u_q(R)$.

If we now collect all the pieces from eqs.~(\ref{tderiv}), (\ref{scalx}) and 
(\ref{supergauss}), with $\mathcal{Q}$ determined from eqs.~(\ref{charge1}) and 
(\ref{charge2}), the action for the collective coordinate $R$ will indeed
be of the form (\ref{504}). The static energy is then given by eq.~(\ref{ECL}) and 
the explicit form of the mass term reads
\begin{align}
\uu(\RR) \equiv \frac{u(R)}{m_A} = \frac{2 \pi}{e^2}\,&\int\limits_0^\infty d\tau
\int\limits_{-\pi}^\pi d\sigma\,\sinh\tau \Bigg\{ \frac{n^2}{2 \RR}\,\eta\,
\mathcal{M}[g] + \RR\,\eta^3\,\mathcal{T}[f] +
\nonumber \\[2mm]
&+ \RR\,f^2\,\eta^3\,q^2 + \frac{\eta}{\RR}\,\left[ (\partial_\sigma q)^2 + 
(\partial_\tau q)^2 \right] - 2 \RR \,\eta^3\,\mathcal{K}[f,g] q\Bigg\}\,.
\label{u}
\end{align}
The explicit form of the local expressions $\mathcal{M}$, $\mathcal{T}$ and 
$\mathcal{K}\equiv \mathcal{Q} / m_A^2$ can be found in appendix \ref{app:mass}. 
To discuss the boundary conditions for $q(s,t)$, it is convenient to rewrite 
eq.~(\ref{u}) in the form
\begin{align}
\uu(\RR) = \frac{2\pi}{e^2}\,\int\limits_0^\infty dt\int\limits_{-\pi}^\pi d\sigma\,
\sinh\tau \Bigg\{ & \frac{n^2}{2 \RR}\,\eta\, \mathcal{M}[g] + 
\RR\,\eta^3\,\Big[ (\deriv f)^2 - 2 q \,\frac{n}{\RR^2}\,\big(\deriv - 2\big)\,
\big[\eta^{-3}\,\partial_\sigma(g\,\eta)\big]\Big] + 
\nonumber \\[2mm]
&+ \frac{n}{\RR}\,\left[ (\partial_\sigma q)^2 + (\partial_\tau q)^2 \right]
+ \RR\, \eta^3\,\Big[ n \,\sin\sigma\,\cosh\tau + q\Big]^2\,f^2 \Bigg\}\,.
\label{ux}
\end{align}
At very small and very large $\tau$, the first line in this equation is suppressed because
either the derivative of the profiles $f$ and $g$, or the profiles themselves vanish. 
The boundary conditions for $q$ are thus determined from the balancing 
of the two terms in the second line of eq.~(\ref{ux}). At $\tau \to 0$, we have
$f \to 1$ by the classical equation of motion, and the last term in 
eq.~(\ref{ux}) dominates. This limit thus yields the estimate 
$q \approx - n \sin\sigma\,\cosh\tau$, which is compatible with the 
Neuman boundary condition
\begin{align}
\frac{\partial q(s,t)}{\partial \tau}\Big|_{\tau \to 0} = 0 
\label{bcq}
\end{align}
also obeyed by the other profiles $f$ and $g$. At $\tau \to \infty$, we have
$f \to 0$ by the boundary conditions and thus the first term in the second line of 
eq.~(\ref{ux}) dominates, which also yields a Neumann boundary condition
\begin{align}
\frac{\partial q(s,t)}{\partial \tau}\Big|_{\tau \to \infty} = 0\,. 
\label{bcq1}
\end{align}
While this result differs from the Dirichlet condition eq.~(\ref{bc1}) obeyed be the other 
profiles, the numerical studies below indicate that $q(s,\infty) \approx 0$ when the 
field equations are obeyed (at the end of the $q$-relaxation).

\subsection{Quantization of the collective coordinate}
\label{collquant}
Up to now, we have only determined the action (\ref{504}) due to the unstable 
mode $R(t)$. To compute the quantum correction to the energy due to this mode, 
we next have to quantize this model. First, we note that the mass term $\uu(\RR)$ 
is coordinate-dependent, so we have to introduce a metric factor $g_{11} = \uu(\RR)$ with 
$g={\rm det}(g)=|\uu(\RR)|$. 
The generalized canonical momentum from the classical action eq.~(\ref{504}) is 
\[
 p \equiv \uu(\RR)\,\frac{d\RR}{d\td}\,.
\]
The coordinate $\RR$ requires an extra connection (Christoffel 
symbol, $\Gamma=\frac{1}{2g}\frac{\partial g_{11}}{\partial\hat{R}}$)
upon quantization
\[
 p = - i \hbar\,\left(\frac{\partial}{\partial \RR} + \frac{1}{2}\,\Gamma \right)
 = - i \hbar \,\left( \frac{\partial}{\partial\RR} + \frac{1}{4}\,\frac{\uu'(\RR)}{\uu(\RR)}
\right)\,.
\]
This quantized momentum obeys the canonical commutation relations $[\RR, p] = i \hbar$ 
and the Hamilton operator of the model (\ref{504}) becomes\footnote{This form of $H$ 
corresponds to a particular operator ordering which \emph{differs} from the
Weyl ordering. Eq.~(\ref{q2}) is instead the standard form in curved space-time, 
which is usually determined by requiring (i) invariance under general coordinate 
transformations and (ii) reduction to the standard Laplace operator in flat space.}
\begin{align}
H = \frac{1}{2}\,g^{-\sfrac{1}{4}}\,p\,g^{\sfrac{1}{2}}\,g^{11}\,p\,g^{-\sfrac{1}{4}} 
+ \EE(\RR) = \frac{\widetilde{p}^2}{2} + \EE(\RR)\,,
\label{q2}
\end{align}
where the transformed momentum is
$\widetilde{p} \equiv \uu^{-\sfrac{1}{4}}\,p\,\uu^{-\sfrac{1}{4}} \,.$

\medskip\noindent
In principle, we would have to find the ground state energy $E_0$ of
the Hamiltonian in eq.~(\ref{q2}), for given background profiles
$f(\sigma,\tau)$ and $g(\sigma,\tau)$ (with $q(\sigma,\tau)$ constructed
from eq.~(\ref{gauss5})), and then minimize $E_0$ rather than eq.~(\ref{ECL}). 
This is \emph{very} expensive computationally, because we need to
solve for $E_0$ at \emph{every} step
of the relaxation, which means solving a variational problem for a highly non-local functional.
As motivated earlier we will follow a simpler approach:
From the definition of $\widetilde{p}$ and $p$, we find the
commutator\footnote{Here and in the rest of this section, we make
$\hbar$ explicit for illustrative purposes.}
\[
\big[\,\widetilde{p}\,,\, \RR\,\big] = - i \hbar\,\uu^{-\sfrac{1}{2}} \,,
\]
and therefore the general uncertainty relation in an arbitrary state $\psi$,
\begin{equation}
 \sigma_\psi(\widetilde{p})^2 \,\sigma_\psi(\RR)^2 \ge \frac{\hbar^2}{4}\,
\left| \big\langle\psi\,|\,\uu^{-\sfrac{1}{2}}\,|\,\psi\big\rangle \right|^2\,.
\label{q3}
\end{equation}
Initially, the Hamiltonian 
and the wave functions are only defined for $\RR > 0$, with a potential singularity
at $\RR=0$ that is avoided self-consistently. Formally, we can
then extend $H$ and $\psi$ to negative $\RR$ in such a way that parity is conserved 
and the ground state $\psi_0$ is symmetric. As a consequence
$\big\langle \,\psi_0 \,\big|\,  \widetilde{p} \,\big| \,\psi_0 \,\big\rangle 
= \big\langle \,\psi_0 \,\big|\, \RR \,\big|\, \psi_0 \,\big\rangle = 0 
$
and eq.~(\ref{q3}) becomes
\begin{equation*}
\big\langle\,\psi_0 \, \big| \, \widetilde{p}^2\, \big|\,\psi_0\, \big\rangle 
\ge \frac{\hbar^2}{4 \,\big\langle \,\psi_0\,\big|\,\RR^2 \,\big|\,\psi_0\,
\big\rangle}\,
\left| \big\langle\psi_0\,|\,\uu^{-\sfrac{1}{2}}\,|\,\psi_0\big\rangle \right|^2\,.
\label{q3a}
\end{equation*}
Next we assume that the ground state $\psi_0$ is localized such that 
$\RR$ is restricted to an effective range $-\RR_0 \lesssim \RR \lesssim \RR_0$,
so that
\[
\big\langle \,\psi_0\,\big|\,\RR^2 \,\big|\,\psi_0\,\big\rangle 
\approx \frac{\RR_0^2}{4} \quad {\rm and} \quad
\left| \big\langle\psi_0\,|\,\uu^{-\sfrac{1}{2}}\,|\,\psi_0\big\rangle \right|^2
\approx u(\RR_0)^{-1} \,.
\]
Using this in the previous equation gives 
\[
\big\langle\,\psi_0\,\big|\,\widetilde{p}^2\,\big|\,\psi_0\,\rangle 
\approx \frac{\hbar^2}{\uu(\RR_0)\,\RR_0^2}\,.
\]
Then the ground state energy is well approximated by
\begin{equation}
\EE_0 = \big\langle\,\psi_0\,\big|\,H\,\big|\,\psi_0\,\big\rangle 
\approx \frac{\hbar^2}{2\,\uu(\RR_0)\,\RR_0^2} + \big\langle\,\psi_0\,
\big|\,\EE(\RR)\,\big|\,\psi_0\,\big\rangle \approx
\frac{\hbar^2}{2\,\uu(\RR_0)\,\RR_0^2} + \EE(\RR_0)\,. 
\label{q4}
\end{equation}
In principle, $\RR_0$ is the radius of the region where
the ground state is localized. For a fixed \emph{externally prescribed} 
radius $\RR$, the ground state will adapt and localize around $\RR_0 \approx \RR$. 
This allows to re-interpret eq.~(\ref{q4}): For an externally prescribed core 
radius $\RR$, the quantum system is localized near $\RR$. 
As with any quantum system, it reacts to the localization with quantum fluctuations that are 
repulsive at very small distances. These fluctuations compete with the classical 
energy $\EE(\RR)$, which favors $\RR \to 0$
according to the instability observed in section~\ref{sec:QS}. These two effects 
balance and the system stabilizes at the minimum of the quantum energy
\begin{equation}
\EE_{q} \approx  \frac{1}{2}\,\frac{\hbar^2}{\uu(\RR)\,\RR^2} + \EE(\RR)\,.
\label{q5}
\end{equation}
This quantum energy differs from the classical energy by the
contribution from the
quantization of the unstable mode. We assume that this contribution 
is the dominant quantum effect in stabilizing
the closed string configuration against shrinking to zero.

\medskip\noindent
The derivation of equation~(\ref{q5}) involved various approximations, but a full quantum 
mechanical treatment is likely to produce something similar
--- maybe with a numerical factor of order one in the quantum correction ---  
because the derived form of the quantum corrected energy is essentially dictated by
fundamental principles of quantum mechanics. For a first estimate and a proof of principle 
of quantum stabilization, eq.~(\ref{q5}) therefore represents 
a completely adequate starting point. 

\subsection{Numerics}
\noindent
For the numerical minimization of the quantum energy in eq.~(\ref{q5}), we employ the relaxation 
technique presented in the last section as follows:  We
\begin{enumerate}
\item[\textbf{1)}] determine the profiles $f$ and $g$ by minimizing the classical 
energy $\EE(\RR)$ from
eq.~(\ref{ECL});
\item[\textbf{2)}] compute the auxiliary field $q$ by minimizing the ($q$-dependent part of the) 
collective mass function $u_q$ in eq.~(\ref{u}) 
(see appendix \ref{app:Q} for explicit 
expressions, in analogy to eq.~(\ref{profdis}) $q_{st}$ is the discretized 
version form of $q(\sigma,\tau)$),
\begin{align}
\frac{\delta \uu}{\delta q_{st}} &= \frac{4 \pi}{e^2}\,
\sinh(\tau_0 + t \Delta_t)\,\Big[ X_{st}\,q_{st} - Y_{st}\Big] \stackrel{!}{=} 0\,;
\label{qrelax}
\end{align}
\item[\textbf{3)}] compute the quantum energy  $\EE_q(\RR)$ from eq.~(\ref{q5}) by
substituting the profiles from~1) and~2).
\end{enumerate}
These operations are performed for each fixed radius $\RR$ separately. Minimization of the 
total quantum energy, $\EE_q$ from eq.~(\ref{q5}), would require us to re-adjust 
the profiles $f$ and $g$ after running through steps \textbf{1)}-\textbf{3)}. This 
procedure would then have to be iterated to convergence since modifying the profiles 
$f$ and $g$ causes $q$ to change as well. Alternatively, we could relax all fields $f$, 
$g$ and $q$ simultaneously using $\uu_q$ as target functional for the minimization of 
$q$ and $\EE_q$ as target functional for $f$ and $g$. In both cases, the minimization 
of $\EE_q$ would be performed with the relaxation condition
\begin{align}
\delta \EE_q = - \frac{\hbar^2}{2 \uu^2\,\RR^2}\,\delta \uu + \delta \EE = 0\,.
\label{relaxax}
\end{align}
Since $\uu$ is at most quadratic in the profiles $f$ and $g$, the variation $\delta \uu$ is 
at most linear. Hence the quantum correction in the relaxation condition (\ref{relaxax}) 
affects the coefficients $H_{st}$, $L_{st}$ (for $f$), and $A_{st}$, $B_{st}$ (for $g$), 
cf.~appendix \ref{app:relax}. Obviously these corrections are proportional to $1/\uu^2$ 
and thus heavily suppressed since our numerical studies below yield $\uu > 100$ for 
all radii at the minimum, and even much larger during relaxation. Therefore the 
impact of the quantum correction on the profiles $f$ and $g$ is a negligible higher 
order effect and the simple procedure \textbf{1)}-\textbf{3)} listed above is
essentially equivalent to the complete minimization of $E_q$ within
our numerical accuracy. 

\begin{figure}[t]
\includegraphics[width = 7cm]{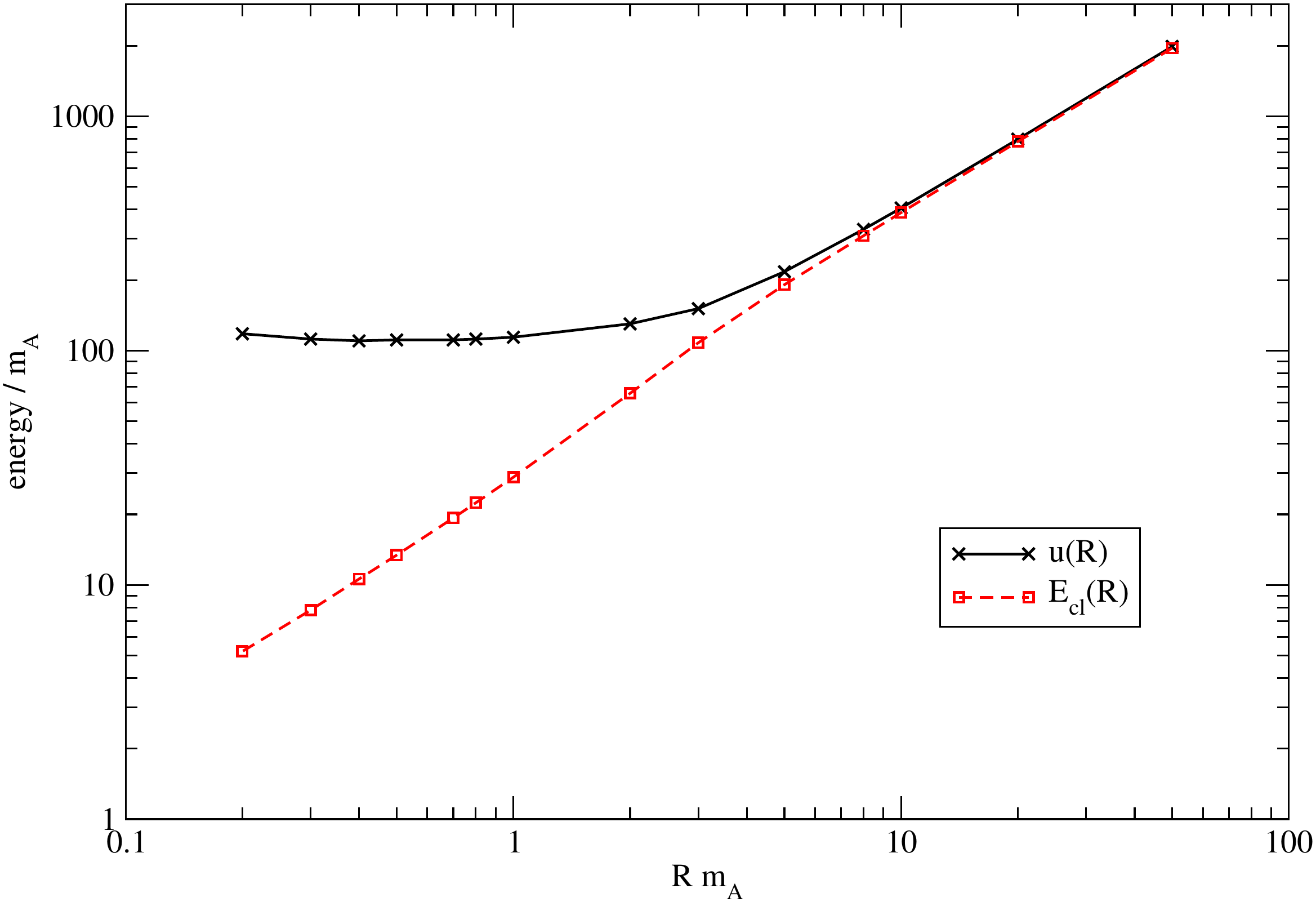} \hspace*{\fill}
\includegraphics[width = 7cm]{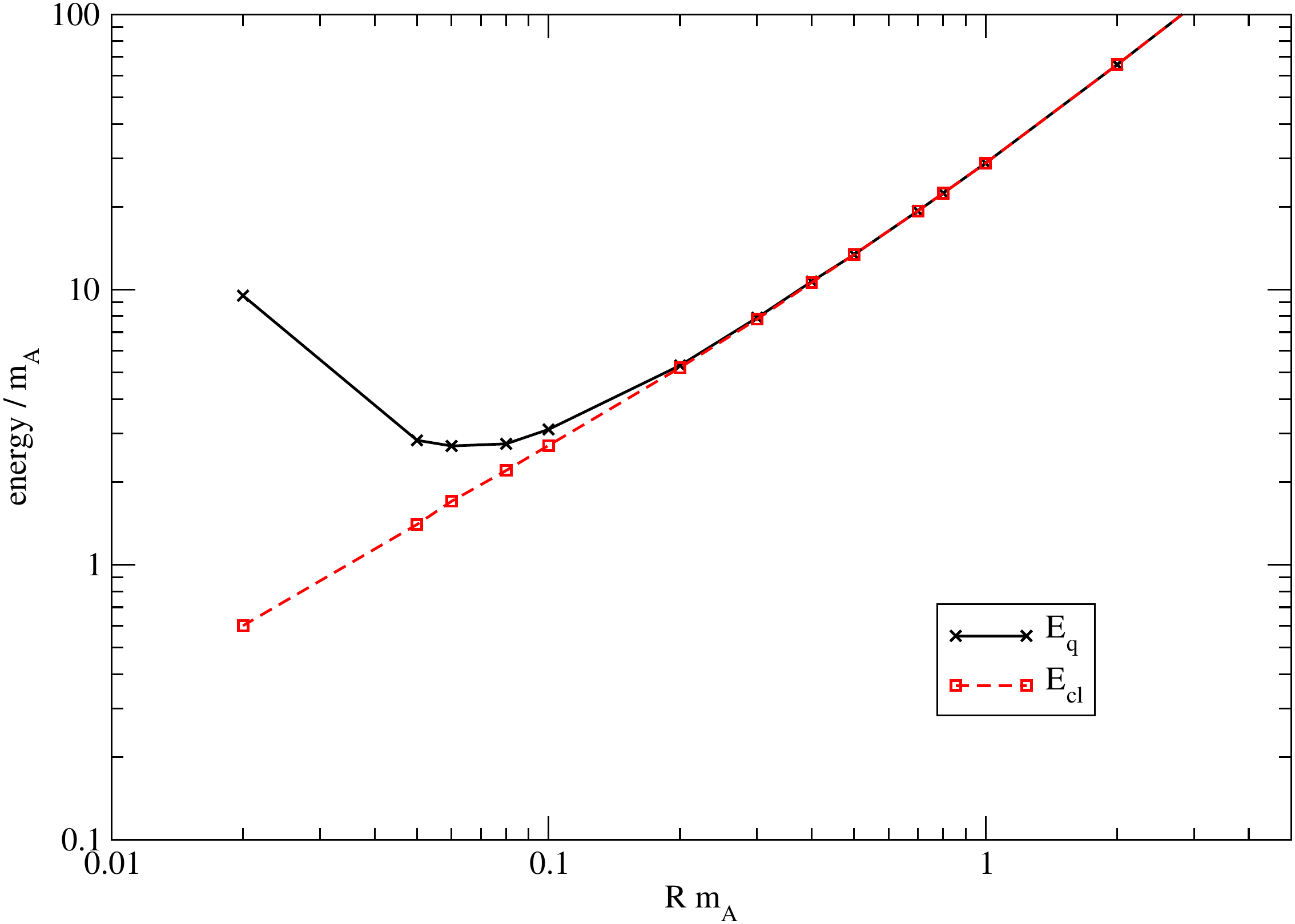}
\caption{(Color online) Left panel: Mass factor of the collective coordinate and classical
energy, as a function of the torus radius. Right panel: Quantum energy vs.~classical 
energy as a function of the torus radius. All simulations were performed with the standard 
parameters $e^2 = 1/2$ and $\beta = 1$.} 
\label{fig:5}
\end{figure}

\medskip\noindent
Figure \ref{fig:5} shows the numerical results for the mass function 
eq.~(\ref{u}) for the collective mode using the profiles obtained according
to steps \textbf{1)} and \textbf{2)}. For large and moderate radii $R$, the mass 
is fairly close to the classical energy, and decreases accordingly when $R$ decreases. 
If this trend continued down to very small radii, we would observe quantum stabilization 
at the expected torus size of a few Compton wave lengths, 
$\RR \approx 1$. However, figure \ref{fig:1} shows that the mass function starts 
to saturate\footnote{It even slightly increases at very small radii, but this could
also be a numerical artifact as the computation becomes delicate when $R \to 0$.} 
at around $\RR \approx 5$ at a large value of $\uu \approx 110$.
As a consequence, the stabilizing radius $\RR \approx 0.1$ in the right panel of figure 
\ref{fig:5} is about one order of magnitude smaller than na{\"\i}vely expected. Although the 
stabilizing mechanism works in principle, the geometry of the stable configuration 
is characterized by such a small radius that it no longer resembles a torus. It is 
suggestive that other quantum effects may have already become relevant for such 
configurations.

A more thorough understanding of the numerical results for $\uu$ is gained by separating 
three contributions in eq.~(\ref{u}): The first two terms are the gauge and Higgs 
kinetic energies, respectively. The third contribution implements Gau\ss{}' 
law.  Table \ref{tab:2} shows some typical results: The gauge kinetic energy 
accounts for about one third of the total mass, while the Higgs kinetic energy and 
Gau\ss{}' law give very large contributions that almost cancel each other. 
In total these two pieces constitute approximately two thirds of the mass. 
On the other hand, if we separate contributions according to eq.~(\ref{ux})
no cancellation of large terms emerges. The reason is that those terms are 
organized as the square of the covariant derivative, $|D_0 \Phi|^2$. 
In that combination Gau{\ss}' law guarantees that there are no large canceling 
contributions to the integral from spatial infinity. If we had not solved 
Gau{\ss}' law, a non-zero Higgs field contribution to the mass function at 
spatial infinity would not be canceled and the mass function would be ill-defined. 
This is reflected by the (almost linear) increase of the Higgs field component 
displayed in table \ref{tab:2}.

\begin{table}[ht]
\centering
\begin{tabular}{|c||c|c|c||c|} \hline
$\RR = R / m_A$ \rule{0mm}{5mm} & \,\,gauge\,\, & \,\,Higgs\,\, & \,\,Gau\ss{}\,\, &
\,\,total\,\, \nonumber 
\\  \hline\hline
$50.0$ &  415.0 & 227075.0 & -225504.0  & 1985.8 \\ \hline
$20.0$ &  167.3 & 90108.0  & -89477.2   & 798.1  \\ \hline
$10.0$ &  86.0  & 44783.1  & -44463.3   & 388.7  \\ \hline
$8.0$  &  70.2  & 35757.6  & -35499.2   & 310.0  \\ \hline
$5.0$  &  47.8  & 22260.0  & -22091.4   & 216.5  \\ \hline
$3.0$  &  36.2  & 13304.9  & -13189.9   & 151.2  \\ \hline
$2.0$  &  33.5  & 8856.5   & -8759.6    & 130.4  \\ \hline
$1.0$  &  32.3  & 4435.2   & -4353.3    & 114.2  \\ \hline
$0.8$  &  32.3  & 3555.9   & -3476.3    & 111.9  \\ \hline
$0.5$  &  32.8  & 2243.4   & -2164.9    & 111.2  \\ \hline
$0.3$  &  31.7  & 1323.4   & -1242.5    & 112.5  \\ \hline
\end{tabular}
\caption{Contributions to the mass function eq.~(\ref{u}) for the collective 
coordinate, as a function of the torus radius. Data was taken on $350\times350$ lattices 
at the standard parameters $e^2 = 1/2$ and $\beta=1$.}  
\label{tab:2}
\end{table}

The accurate determination of the mass factor eq.~(\ref{u}) requires a high resolution 
in the (tiny) region between the torus core at $R$ and the origin. In addition, the 
overall grid extension must be large enough to reach asymptotic distances. 
The numerical parameter $d_{\rm max}$, which controls both the extension towards 
spatial infinity and the distance from the symmetry axis, is measured in units of 
$R$. On the other hand, the physical length scale is the Compton wave length, $1/m_A$.
Hence $d_{\rm max}$ has to be very large, especially when $\RR < 1$. In practice, 
values of $d_{\rm max} \simeq 200$ and grid sizes up to $500\times500$ were necessary 
to obtain stable results. 

For completeness, we have plotted the induced profile function $q$, that emerges from 
Gau{\ss}' law for two different radii in figure \ref{fig:q}. Deviations from the vacuum 
value $q=0$ extend much further out from the core region than the profiles $f$ and $g$, 
cf.~figure \ref{fig:new}. Also, the main structure is located above and below the plane 
of the torus (core line) and $q$ changes sign across this plane. The $q$-field vanishes 
in the core plane, which is not implemented as a boundary condition, cf. eq.~(\ref{bcq1}), 
but rather a result obtained when the relaxation has converged and thus it is a consequence 
of Gau\ss{}' law. 

For very small radii, rather large values of $d_{\rm max}$ are required in order to cover 
the region near the vertical axis in figure \ref{fig:q}, in which the deviation from the 
vacuum is sizeable.  For illustration purposes $d_{\rm max}$ has been chosen fairly small 
($d_{\rm max} \approx 10$) in figure \ref{fig:q} while in the actual numerical calculations 
it was taken at least an order of magnitude larger.

\begin{figure}[t]
\includegraphics[width=8.0cm,height=7.5cm]{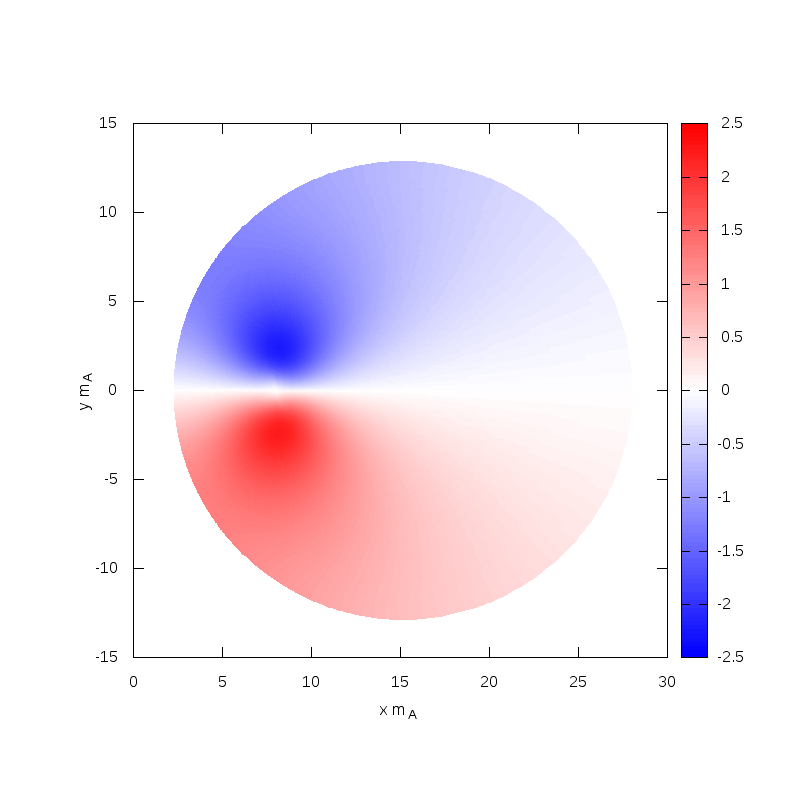} \hspace*{\fill}
\includegraphics[width=8.0cm,height=7.5cm]{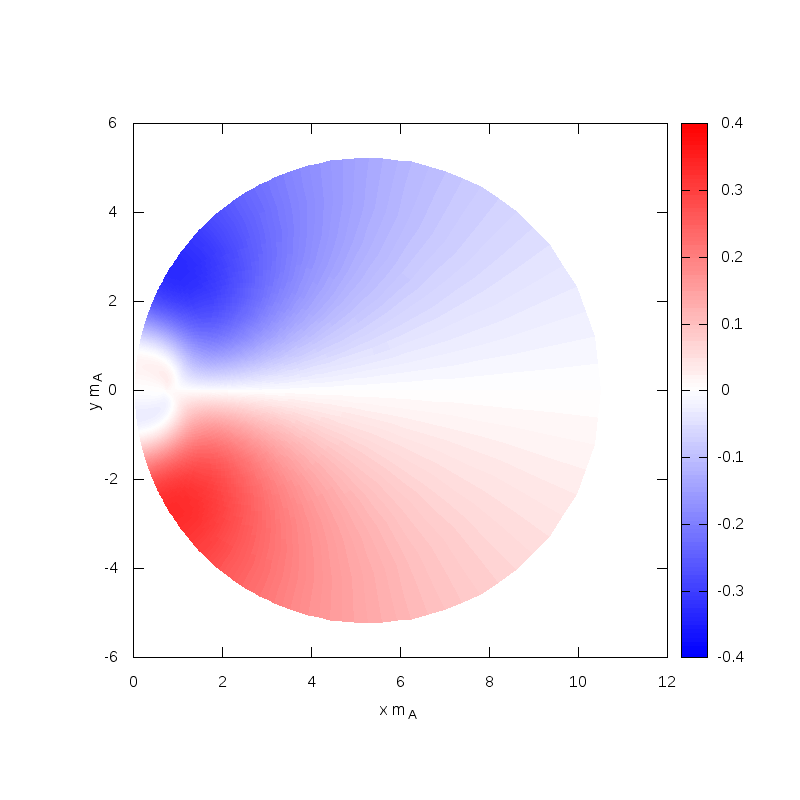}
\caption{(Color online) Density plots of the profile function $q$ in a plane 
perpendicular to the torus.  In the left panel, the torus has a radius 
$\RR=8$, while in the right panel shows results for $\RR=0.7$. Notice the 
different scales on the axes in both cases. The profile is positive (negative) 
in the lower (upper) half-plane.}
\label{fig:q}
\end{figure}

\section{Conclusion}

The Nielsen-Olesen string in scalar electrodynamics is a suitable
toy model for cosmic strings in non--Abelian gauge theories such as 
the Standard Model. For this
reason we have investigated the stability issue of the closed 
Nielsen-Olesen string. The closed string has been constructed by
introducing toroidal coordinates. This complicated matters significantly
because the profiles become functions of two independent variables,
and quantitative analyses cannot be pursued in from of ordinary 
differential equations. Therefore we adopted a lattice approach 
and performed variational calculations by relaxation methods. 
This verified the classical instability of a toroidal string as the
closed string shrinks to size zero.  As in the case of the classical
hydrogen atom, by radiating energy as fluctuations and simultaneously
shrinking, the closed string will eventually decay. 

In the context of the cosmological assumption that large scale closed 
cosmic strings could have existed at early times, this analysis suggests that 
(non-interacting) closed strings would not have survived. We have then 
argued that quantum mechanics in form of the uncertainty principle 
disallows objects that at some stage possessed a finite size squeeze to
arbitrarily small extensions. Motivated by this observation we have developed 
a formalism for the quantum stabilization. It is based on the uncertainty 
principle and inspired by its success in reproducing the fundamental 
quantum mechanical properties of the hydrogen atom.

Technically this approach requires the computation of an inertia function 
for a time dependent radius of the string. This inertia function can be 
interpreted as a mass for the radius of the torus and we have estimated its 
kinetic energy term in the quantum energy functional. We have 
then applied a variational principle to this functional (appropriately 
extending the configuration manifold properly to implement Gau{\ss}' law) 
and determined the field configuration from the minimal quantum energy at 
a prescribed radius. As expected from the uncertainty principle, the kinetic 
energy part is negligible for large radii. More surprisingly, it is also quite 
small for radii that are of the scale of the Compton wave-length of the 
fundamental fields. Nevertheless we have observed that the proposed mechanism of 
quantum stabilization is indeed operative in the case of the closed Nielsen-Olesen 
string as the quantum energy exhibits a global minimum for a non-zero radius. 
The proper incorporation of Gauss' law proved essential for this
mechanism to work. However, the numerical results for the radius at at 
which the torus becomes stable suggest that other mechanisms should be
relevant as well, since this radius is about an order of magnitude
smaller than the length-scale inherent in the model. (We do not expect 
the corrections to the approximations in section~\ref{collquant} to compensate 
for this.) At such small radii the toroidal character of the shrinking 
configuration has disappeared and presumably additional quantum modes (besides 
the length scale) must be included when computing the quantum
energy. This conclusion is 
further supported by the fact that the mass associated with the radial mode is quite 
large. Hence the spectrum of radial excitations will be very dense. Presumably 
these additional quantum fluctuations will be enhanced in the interior region of 
the string. Rather than decaying to infinity, these fluctuations would increase the 
energy of the string thereby stabilizing it at radii larger than those computed here. 
This suggests that the stabilization radius obtained from the application 
of the uncertainty principle to the radius of the closed string is merely a lower 
limit. Of course, due to the complicated structure of the background configuration 
any quantitative analysis of this scenario will be a formidable task.

\begin{acknowledgments}
N.\ G.\ was supported in part by the National Science Foundation (NSF)
through grant PHY-1213456.
H.\ W.\ is supported by NRF (Ref. No. IFR1202170025).
\end{acknowledgments}

\section*{Appendices}
In these appendices we present details of the numerical
approach used in sections \ref{sec:toro} and \ref{sec:QS}.

\appendix
\subsection{Lattice discretization}
\label{app:lattice}
\noindent
We start with the discretized description eq.~(\ref{lat1}), where
the profiles turn into real functions 
\begin{align}
f_{st} \equiv f(s\,\Delta_s, \tau_0 + t\,\Delta_t)\,,\qquad\qquad
g_{st} \equiv g(s\,\Delta_s, \tau_0 + t\,\Delta_t)
\label{profdis_app}
\end{align}
defined on the lattice sites. The boundaries of the parameter regions are at 
$t=0$ and $t=N_t$, and at  $s=0$ and $s=N_s$, respectively, where we impose 
the following conditions,
cf.~eqs.~(\ref{bc1})--(\ref{bc3}):
\begin{itemize}
\item periodic boundary conditions on $s$:
\begin{equation}
f_{0t} = f_{N_st}\,,\qquad\qquad g_{0t} = g_{N_s t}\,,\qquad\quad
t \in \{0,\ldots,N_t\}
\label{bcx1}
\end{equation}
\item Dirichlet conditions at $t=N_t$:
\begin{equation}
f_{s N_t} = 0\,\qquad\qquad g_{s N_t} = 0\,,\qquad\qquad\quad
s \in \{ 0,\ldots\,N_s\}
\label{bcx2}
\end{equation}
\item Neumann conditions at $t=0$:
\begin{equation}
f_{s0} = f_{s1}\,,\qquad\qquad g_{s0} = g_{s1}\,,\qquad\qquad 
s \in \{0,\ldots,N_s\}\,.
\label{bcx3}
\end{equation}
\end{itemize}

\noindent
Notice that the derivatives in the Neumann conditions are \emph{forward}
derivatives at $t=0$ so that no extra grid points are required. For integrals over 
$s$ and $t$, we use the same prescription combined with a 
\emph{left} Riemann sum definition,
\[
\int\limits_{\tau_0}^{\tau_\infty} d\tau \,u(\tau) 
\to \sum_{t=0}^{N_t-1} \Delta_t\,u(\tau_0 + t\,\Delta_\tau)\,.
\]
This ensures that we will 
never encounter grid points outside the range defined in eq.~(\ref{lat1}).
Next, we choose the discretized profiles $f_{st}$ and $g_{st}$ such as to minimize the 
(discretized) target functional eq.~(\ref{ECL}),
\begin{eqnarray}
\EE &=& \frac{\pi}{e^2}\sum_{t=0}^{N_t-1}\sum_{s=0}^{N_s-1}\Delta_t
\Delta_s\,\sinh(\tau_0 + t \Delta_t)\,\Bigg\{ \frac{n^2}{\RR}\,
\frac{1}{\eta_{st}}\,\left[\frac{g_{s,t+1}- g_{st}}{\Delta_t}\right]^2 + 
\label{targetx}\\[2mm]
&& {}+ \RR\,\eta_{st}\,\left[ \left(\frac{f_{s,t+1}-f_{st}}{\Delta_t}
\right)^2 + n^2 f_{st}^2 (1 - g_{st})^2+ \left(\frac{f_{s+1,t}-f_{st}}{\Delta_s}
\right)^2 \right] +  
\frac{\beta}{4}\,\RR^3\,\eta_{st}^3\,(1-f_{st}^2)^2 \Bigg\}\,,
\nonumber
\end{eqnarray}
where the dimensionless metric factor is given in eq.~(\ref{dimparm}).
Once the solutions $f_{st}$ and $g_{st}$ are found, the original profiles 
$f(\vek{x})$ and $g(\vek{x})$ can be restored from eq.~(\ref{profdis_app})
and the definition of toroidal coordinates, eq.~(\ref{toro}).

\subsection{Relaxation}
\label{app:relax}
\noindent
For the relaxation condition, we consider an interior point $(s,t)$ at which 
the profiles are not fixed by boundary conditions. At these points we can 
independently vary the profile functions and compute the associated 
gradients. We start by listing the result for the gauge field:
\begin{equation}
\frac{\delta \EE}{\delta g_{st}}=
\frac{2\pi}{e^2} n^2 \,\sinh(\tau_0 + t \Delta_t)\,
\Big[ A_{st} \,g_{st} - \,B_{st}  \Big]
\label{relax1}
\end{equation}
with
\begin{eqnarray}
A_{st} &=& \frac{\varepsilon}{\RR}\,\left( \frac{1}{\eta_{st}} +
\frac{\xi_{t}}{\eta_{s,t-1}} \right) + \RR\,f_{st}^2\,\eta_{st}\,\Delta^2\,,
\nonumber \\[2mm]
B_{st} &=& \frac{\varepsilon}{\RR}\,\left(\frac{g_{s,t+1}}{\eta_{st}}
+ \xi_{t}\,\frac{g_{s,t-1}}{\eta_{s,t-1}} \right) +
\RR\,f_{st}^2\,\eta_{st}\,\Delta^2\,.
\nonumber
\end{eqnarray}
For later convenience, we have introduced the abbreviations
\begin{eqnarray}
\xi_{t} &\equiv& \frac{\sinh(\tau_0 + (t-1)\,\Delta_t)}{\sinh(\tau_0 + t \Delta_t)}
= \cosh(\Delta_t) - \coth(\tau_0 + t \Delta_t)\,\sinh(\Delta_t)\,,
\nonumber \\[2mm]
\varepsilon &\equiv& \frac{\Delta_s}{\Delta_t} 
\qquad {\rm and} \qquad
\Delta^2 \equiv \Delta_s\,\Delta_t \,.\nonumber 
\end{eqnarray}
The local minimization of $\EE$ with respect to
$g_{st}$  yields the updating rule (\emph{relaxation condition})
\begin{equation}
g_{st} \longrightarrow \frac{B_{st}}{A_{st}}\,,
\label{relax2}
\end{equation}
which in view of $A_{st} > 0$ is a minimum.
For the Higgs profile, $f_{st}$, a similar calculation gives
\begin{equation}
\frac{\delta \EE}{\delta f_{st}}=
\frac{2 \pi}{e^2}\,\sinh(\tau_0 + t \Delta_t)\,\hat{R}\,
\Big[ 2 G_{st}\,f_{st}^3 + H_{st}\,f_{st} - L_{st} \Big]\,,
\label{relax3}
\end{equation}
with the coefficients
\begin{eqnarray}
G_{st} &=& \frac{\beta}{4}\,\RR^2\,\eta_{st}^3\,\Delta^2\,,
\nonumber \\[2mm]
H_{st} &=& \varepsilon\,\left(\eta_{st} + \eta_{s,t-1}\,\xi_{t}\right) + 
\frac{1}{\varepsilon}\,\left(\eta_{st} + \eta_{s-1,t}\right) 
+ n^2 \,\eta_{st}\,(1 - g_{st})^2 \,\Delta^2 - \frac{\beta}{2}\, 
\RR^2\,\eta_{st}^3\,\Delta^2\,,
\nonumber \\[2mm]
L_{st} &=& \varepsilon\left( f_{s,t+1}\,\eta_{st} 
+ f_{s,t-1}\,\eta_{s,t-1}\,\xi_{t}\right) + 
\frac{1}{\varepsilon}\left( f_{s+1,t}\,\eta_{st} + 
f_{s-1,t}\,\eta_{s-1,t}\right)\,.
\nonumber
\end{eqnarray}
These values are such that the \emph{relaxation condition}, 
$\frac{\delta \EE}{\delta f_{st}}=0$, always has a real solution
$f_{st}^{(0)}$ which can be computed by a closed formula
into which we substitute the numerically obtained data for the 
coefficients $G_{st}$, $H_{st}$ and $L_{st}$.  
The updating rule for the Higgs field is then 
\begin{equation}
f_{st} \longrightarrow f_{st}^{(0)}\,.
\label{relax4}
\end{equation}
Since the relaxation conditions (\ref{relax2}) 
and (\ref{relax4}) only compute a local minimum with all other field variables 
held fixed, we must proceed iteratively and sweep through the entire 
lattice many times to relax to an overall stable minimum.

\subsection{Mass factor for the collective coordinate}
\label{app:mass}
The action for the collective coordinate $R$, eq.~(\ref{504}), contains a mass 
factor $u$ that is a local functional of the profiles. The three abbreviations
in the explicit form eq.~(\ref{u}) read 
\begin{align}
\mathcal{M} &=  \big( \cos(2\sigma) + \cosh(2 \tau) \big)\,g^2 + 
4 \,\cos\sigma\,\cosh\tau\, g \deriv g 
+ 2 \,\big(\deriv g\big)^2
\nonumber \\[2mm]
\mathcal{T} & = (\deriv f)^2 + \left(n\,f\,\sin\sigma\,\cosh\tau\right)^2
\nonumber \\[2mm]
\mathcal{K} &= - n \,f^2\,\sin\sigma\,\cosh\tau + \frac{n}{\RR^2}\,
\big(\deriv - 2\big)\,\left[ \eta^{-3}\,\partial_\sigma(g\,\eta)\right] \,.
\label{uabbrev}  
\end{align}
Note that the profile functions $f$ and $g$ depend on the toroidal
coordinates $\sigma$ and $\tau$. They also have an implicit dependence on
the torus radius $R$.

\subsection{Quantum relaxation condition}
\label{app:Q}
The relaxation condition for the auxiliary field $q_{st}$ related to Gau\ss{}' law
was given in eq.~(\ref{qrelax}) in the main text. The corresponding coefficients read 
explicitly
\begin{align}
X_{st} &=  \RR\, \eta_{st}^3\,f_{st}^2\,\Delta^2 + \frac{1}{\RR}\,
\Big[ \xi_t\,\epsilon\,\eta_{s,t-1}\, + \frac{1}{\epsilon}\,\eta_{s-1,t} + 
\left(\epsilon + \frac{1}{\epsilon}\right)\,\eta_{st} \Big]
\nonumber
\\[2mm]
Y_{st} &= \RR\,\eta_{st}^3\,\mathcal{K}_{st}\,\Delta^2 + \frac{1}{\RR}\,\Big[
\epsilon\,\xi_t\,\eta_{s,t-1}\,q_{s,t-1} + \frac{1}{\epsilon}\,\eta_{s-1,t}\,q_{s-1,t}
+ \epsilon\,\eta_{st}\,q_{s,t+1} + \frac{1}{\epsilon}\,\eta_{st}\,q_{s+1,t}
\Big]\,,
\label{qcoeff}
\end{align}
where $\mathcal{K}_{st}$ is the descretized (lattice) version of $\mathcal{K}$
in eq.~(\ref{uabbrev}).

\end{document}